\documentclass[final,5p,times,twocolumn]{elsarticle}

\usepackage{lineno}
\usepackage{longtable,supertabular}
\usepackage{enumitem}

\biboptions{sort&compress}

\usepackage{esvect}
\usepackage{textgreek}
\usepackage{textcomp}
\usepackage{amssymb}
\usepackage[export]{adjustbox}
\usepackage{subcaption}
\usepackage{caption}
\usepackage{wrapfig}
\usepackage{graphicx}
\usepackage{lipsum}
\usepackage{float}
\usepackage{bm}
\usepackage{makecell}
\usepackage{siunitx}
\usepackage{amsmath}
\usepackage{xcolor}
\usepackage{tabularx}
\usepackage{tikz}
\journal{Acta Materialia}

\usepackage{multirow}
\graphicspath{{Figures/}{SuppFigures/}}

\newcommand{\numGBs}{7272 }
\newcommand{\numAtoms}{70 million }

\newcommand{\documenttitle}{Grain boundary solute segregation across the 5D space of crystallographic character}

\begin{document}
\begin{frontmatter}

\title{\documenttitle}

\author[label1,label3]{Lydia Harris Serafin}
\author[label2]{Ethan R. Cluff}
\author[label1]{Gus L. W. Hart}
\cortext[cor1]{Corresponding author}
\author[label2]{Eric R. Homer\corref{cor1}}
\ead{eric.homer@byu.edu}

\address[label1]{Department of Physics and Astronomy, Brigham Young University, Provo, UT, 84602, USA.}
\address[label2]{Department of Mechanical Engineering, Brigham Young University, Provo, UT, 84602, USA.}
\address[label3]{X-Computational Physics Division, Los Alamos National Laboratory, Los Alamos, NM, 87545, USA.}

\begin{abstract}

Solute segregation in materials with grain boundaries (GBs) has emerged as a popular method to thermodynamically stabilize nanocrystalline structures. However, the impact of varied GB crystallographic character on solute segregation has never been thoroughly examined. This work examines Co solute segregation in a dataset of \numGBs Al bicrystal GBs that span the 5D space of GB crystallographic character. Considerable attention is paid to verification of the calculations in the diverse and large set of GBs. In addition, the results of this work are favorably validated against similar bicrystal and polycrystal simulations. As with other work, we show that 
Co atoms exhibit strong segregation to sites in Al GBs and that segregation correlates strongly with GB energy and GB excess volume. Segregation varies smoothly in the 5D crystallographic space but has a complex landscape without an obvious functional form.
\end{abstract}


\end{frontmatter}



\section{Introduction} \label{sec:intro}

Solute atoms in polycrystalline materials with grain boundaries (GBs) may stay in the bulk, diffuse to the surface, or segregate to the GB, among other behaviors such as forming precipitates. When solute atoms segregate, they often remain in the GB network due to both kinetic mechanisms and thermodynamic stabilization. Some examples of kinetic mechanisms are solute drag that slows GB mobility \cite{Humphreys:2004tg} and solute pinning that prevents GB mobility under external driving forces \cite{lucke1957quantitative}. Thermodynamic stabilization involves lowering the Gibbs free energy of a GB interface by the presence of the solute atom \cite{liu2004nano, kalidindi2017stability}, and is described in a theoretical framework developed by Weissm\"uller \cite{weissmuller1994alloy, weissmuller1994some}. Thermodynamic stabilization can be utilized to engineer materials with greater hardness than a pure material even at elevated temperatures (e.g., in \cite{liu2004nano, Ruan:2009kq, Chookajorn:2012fo, shi2023effects}) due to the Hall-Petch effect that causes greater hardness with smaller grain sizes \cite{Hall:1951cy, Petch:1953ws}. A recent review of thermodynamic stabilization is given in \cite{perrin2021stabilized}.

Early on, simple effective segregation energy models of solute segregation \cite{mclean1957isotherm} were derived from experimentally determined values and are still often used to predict solute concentration at GBs and in larger mesoscale models \cite{friedel1954electronic, seah1980adsorption, erhart1981equilibrium, muschik1989surface, christien2003phosphorus, robson2014effect}.\footnote{Other models of segregation energy are reviewed in Chapter 4 of \cite{Lejcek:2010un}.} However, such models violate the third law of thermodynamics \cite{wagih2021thermodynamics}, don't account for the effects of GB character on the system, and are insufficient to describe experimental behavior \cite{white1977spectrum, mutschele1987segregation, kirchheim1988hydrogen}, in particular of strong segregation to specific atom sites  (e.g., in the Co-Al-W system \cite{titus2016solute}).

Recently, models that are informed by simulated data have been developed to address such issues. GB solute segregation has been examined in the dilute limit in atomistic studies utilizing polycrystals \cite{wagih2019spectrum, wagih2020learning, barr2021role, wagih2023spectrum, ito2023analysis, tuchinda2024computed} and bicrystals in small regions of the 5D GB space \cite{huber2018machine, mahmood2022atomistic, cui2023effect, zhang2023weight, menon2024atomistic}, as well as in experimental studies \cite{pei2023atomistic, barr2021role}. Some studies improve on the segregation energy calculations by including entropic and other effects in the atomistic simulations \cite{tuchinda2024computed, wagih2024grain}. Others move beyond the dilute limit by considering solute-solute interactions \cite{zhang2024grain}. From these data, segregation models are often created using machine learning techniques \cite{huber2018machine, wagih2020learning, mahmood2022atomistic}. Recent reviews of computational modeling of solute segregation are given in \cite{lejvcek2017interfacial,hu2024computational}.
Here, we highlight two recent and notable efforts to create more accurate segregation energy models based on segregation energy spectra, which permits enforcement of the third law of thermodynamics \cite{wagih2021thermodynamics}.

Huber et al.\ created a small dataset of densely sampled $\Sigma5$ coincident site lattice (CSL) Al bicrystal GBs and calculated segregation energy spectra for 6 different solute types to inform machine learning models for solute segregation at the atomic level \cite{huber2018machine}. They found that a thorough sampling of the 5D space of GB crystallographic character was necessary for the creation of a segregation energy spectrum that informs a model more accurate than the rudimentary effective segregation energy model. 
Wagih et al.\ \cite{wagih2023spectrum} also point out issues with using an effective segregation energy for solute concentration models, demonstrated by the bimodal spectrum of segregation energies in the Pd-H system due to the occupation of interstitial sites.
In response, they created solute segregation spectra in polycrystals for use in segregation energy models by performing atomistic simulations of Mg in Al polycrystals \cite{wagih2019spectrum}. This polycrystal approach creates one segregation energy spectrum for the entire dataset, rather than many small spectra that are concatenated to represent the dataset, as is necessary for bicrystal GBs. Wagih et al.\ used machine learning on these polycrystal spectra to inform 259 binary alloy system segregation models \cite{wagih2020learning}, as well as some quantum-accurate models \cite{wagih2022learning}. 

Wagih et al.\ use polycrystal simulations in order to more fully capture the behavior of real materials, and they caution against using bicrystal simulations \cite{wagih2019spectrum}. They suggest that only thorough samplings of bicrystal GBs in the 5D space of crystallographic character should be used for this purpose, promoting the work by Huber et al.\ in \cite{huber2018machine} as an example of sufficient sampling, albeit in a small subspace of the 5D space. Others have also noted the insufficiency of bicrystal GB simulations, such as Tucker et al.\ who use the strain functional description of atomic configurations to show that symmetric twist GBs (STGBs) cannot be used to represent polycrystals or amorphous structures \cite{Tucker2022}. Wagih et al.\ also present evidence that STGBs and low coincidence site lattice (CSL) GBs do not represent polycrystals generally \cite{wagih2023can}. 


In this work, we compute the segregation energy of \numAtoms Co atoms in \numGBs Al bicrystal GBs from the Homer GB dataset \cite{Homer:2022:AlGBdataset}. The use of this dataset attempts to address most of the concerns raised about using bicrystal GBs simulations to inform GB solute segregation models in \cite{wagih2023can, Tucker2022, wagih2019spectrum} because it spans the 5D space of GB crystallographic character. It is also not limited to STGBs or low CSL GBs; it includes CSL values up to $\Sigma 999$. Additionally, the use of this dataset is a step towards examining the behavior of solute segregation across the entire GB space, which is noted as an important next step for the field \cite{hu2024computational}. We describe the methods for simulating and computing segregation energies and solute concentrations. We then verify and validate the resulting data, including direct comparisons to the works by Huber et al.\ in \cite{huber2018machine} and Wagih et al.\ in \cite{wagih2020learning}. 
Finally, we examine the spectra using several techniques, including a statistical overview, a classification scheme, and reduction to a solute concentration for each GB, and we identify some subsets of GBs that deviate significantly from the mean solute concentration.

\section{Methods}\label{sec:methods}

\subsection{Theory of solute segregation}

The segregation energy $E_{\mathrm{seg}}$ of an atom is defined as the energy difference between a solute atom and a solvent atom at the same site in the GB minus the same energy difference at a reference atom site located in bulk \cite{Lejcek:2010un, wagih2019spectrum}. In this work, we examine the segregation of Co in Al, which is calculated according to: 
\begin{equation} \label{eq:segEng}
    E_{\mathrm{seg}}^{\mathrm{Co}_i} = \left(E_\mathrm{Co}^{i} - E_\mathrm{Al}^{i}\right) - \left(E_\mathrm{Co}^\mathrm{ref} - E_\mathrm{Al}^\mathrm{ref}\right)
\end{equation}
where $E_{\mathrm{Co}}^{i}$ is the energy of a Co atom in the \textit{i}-th atomic site in a GB, $E_\mathrm{Al}^{i}$ is the energy of an Al atom in the \textit{i}-th atomic site, and $E_\mathrm{Co}^\mathrm{ref}$ and $E_\mathrm{Al}^\mathrm{ref}$ are the energies for a reference atom in bulk, far away from the GB. All of these values are calculated at $0$ K. In this formulation, segregation is energetically favorable for a site when $E_{\mathrm{seg}}^{\mathrm{Co}_i}$ is negative.

\subsection{Solute segregation energy spectrum creation}

In this work, segregation energy data is collected by substituting single Co atoms into Al GBs from a dataset created by Homer et al.\ \cite{Homer:2022,Homer:2022:AlGBdataset} which used the pure Al EAM potential from Mishin et al.\ \cite{Mishin:1999tg}. This dataset is referred to in the present work as the Homer dataset. The Homer dataset contains GB structures that have 150 different CSL values corresponding to unique disorientations up to $\Sigma 999$, sampled at intervals of \mbox{$\sim$$5^{\circ}$} in the disorientation space. For each CSL value, a sampling of boundary planes (BPs) was selected to provide comprehensive coverage, making 7304 unique GBs in the 5D space of GB crystallographic character.\footnote{10 of the 7304 GB structures in the Homer dataset \cite{Homer:2022:AlGBdataset} are excluded from this work due to computational difficulties. See Supplemental Table \ref{tab:excludedGBs} for a list of excluded GBs.} The optimal atomic configuration for each GB was then found by varying 6 parameters of GB construction while maintaining the 5D constraint of the GB, relaxing each structure via conjugate gradient energy minimization. In this work we examine only the minimum energy configuration of the 6 GB construction parameters. 
See \cite{Homer:2022} for additional details about the construction of the Homer dataset.

Segregation energy values in Equation \ref{eq:segEng} were computed in LAMMPS molecular statics simulations \cite{LAMMPS} using the Ni-Al-Co empirical EAM potential from Pun et al.\ \cite{pun2015interatomic}, which is the same potential used by Huber et al.\ in \cite{huber2018machine} and Wagih et al.\ in \cite{wagih2020learning}. We start with a relaxed GB structure and replace an existing Al atom with a Co atom at the same atom site. The entire GB structure is then relaxed, and the segregation energy for the substituted atom, $E_\mathrm{seg}^{\mathrm{Co}_i}$, is calculated using Equation \ref{eq:segEng}.

This process was completed for approximately \numAtoms
atoms from \numGBs GBs: each atom closer than \mbox{15 \AA} to the GB plane was replaced by a Co atom, as well as a random sample of 100 atoms for each GB in the range of \mbox{$15\text{--}25$ \AA}\ to use for bulk reference energies \mbox{($E_\mathrm{Co}^\mathrm{ref}$ \& $E_\mathrm{Al}^\mathrm{ref}$)}.  
Atoms further away from the GB are not substituted because $E_\mathrm{seg}^{\mathrm{Co}_i}$ rapidly falls to \mbox{$0$ eV} with distance from the GB plane \cite{jin2014study, scheiber2015ab}.  

\subsection{Analysis techniques}
\label{sec:methods-analysis}

Since the segregation data for each GB results in a spectrum of values, we employ two methods to simplify comparison of the spectra across the set of GBs. Specifically, these methods are i) a classification scheme and ii) a grain boundary solute concentration, which are described below.

For reasons that will be clear in the verification section (\ref{sec:bulkSeg}), we implement a classification scheme that classifies any atom with near-bulk segregation behavior as ``negligibly segregating.'' We do this because the segregation energy values of the bulk atoms actually take on a range of values
about \mbox{$0$ eV} and GB atoms with segregation energy values in that same range would behave the same as if they were in bulk. We designate segregation energy values in the 95\% interval\footnote{Supplemental Figure \ref{fig:neglLimitsFCC} shows a number of intervals on the distribution of FCC atoms, from which we determined to use a 95\% interval for the ``negligible'' classification. Supplemental Figure \ref{fig:neglLimitsNonFCC} shows a number of intervals on the distribution of non-FCC atoms.} of the bulk atom distribution as ``negligibly segregating''. This allows us to more easily determine which atoms are ``segregating'' and ``anti-segregating'' because they are outside the range of typical bulk atom segregation energies. The ranges for these three possible classifications are: i) segregating; \mbox{$E_{\mathrm{seg}}^{\mathrm{Co}_i} < -0.0875$ eV}, i) negligibly segregating; \mbox{$-0.0875$ eV $\leq E_{\mathrm{seg}}^{\mathrm{Co}_i} < +0.018$ eV}, and i) anti-segregating; \mbox{$E_{\mathrm{seg}}^{\mathrm{Co}_i} \geq +0.018$ eV}. 

A standard measure of solute segregation at GBs is the solute concentration at the GB in the dilute limit, $c_\mathrm{GB}$. Early literature calculated this value using a single effective segregation energy \cite{mclean1957isotherm} or a continuous distribution of segregation energies for atoms in the GB \cite{seah1973grain}. This type of approach is computationally simple but has some pitfalls (as mentioned in Section \ref{sec:intro}) that can be avoided by using a discrete segregation energy spectrum. Coghlan and White first created such a spectrum \cite{white1977spectrum}, which was later adapted by Huber et al.\ \cite{huber2018machine} for an array of individually described atoms in a single GB. The concentration of solute atoms in the GB is then calculated according to:

\begin{equation} \label{eq:conc}
    c_{\mathrm{GB}} = \frac{1}{N} \sum_i \left[1 + \frac{1 - c_{\mathrm{bulk}}}{c_\mathrm{bulk}} \mathrm{exp}\left(E_{\mathrm{seg}}^{\mathrm{Co}_i}/k_\mathrm{B} T \right)\right]^{-1}
\end{equation} 

\noindent where $E_{\mathrm{seg}}^{\mathrm{Co}_i}$ is defined by Equation \ref{eq:segEng}, $N$ is the number of sites in the GB, $k_\mathrm{B}$ is the Boltzmann constant, $T$ is the temperature, and $c_{\mathrm{bulk}}$ is the concentration of solute in bulk, held fixed as an independent variable. Bulk atom sites are chosen for solute occupancy at finite temperatures with increasing probability, lowering $c_\mathrm{GB}$; the temperature dependence of this value is demonstrated in Section \ref{sec:results}. The $\frac{1 - c_{\mathrm{bulk}}}{c_\mathrm{bulk}}$ term scales the Fermi level down as bulk atom sites are filled at finite temperatures, as discussed in \cite{huber2018machine}. In this work we use a bulk concentration of \mbox{$c_\mathrm{bulk} =0.2$at\%}, chosen to be the same as in Huber et al.\ \cite{huber2018machine}.

\section{Verification and Validation} \label{sec:VandV}

An important step in collection of any data is the verification and validation of the results \cite{2019.TMS.VandV}. In the following sections we verify that the calculated values are representative of true segregation energies and validate the results by comparing them to other published examples.

\subsection{Verification} 
In verifying the data collected in this work, we noted that some data was incorrect or did not match expected behavior. The following sections discuss the process of determining which data could be verified for accuracy and inclusion in the work.

\subsubsection{Verification of bulk segregation energies} \label{sec:bulkSeg}

By definition (Equation \ref{eq:segEng}), solute atoms in the bulk have segregation energies of \mbox{$0$ eV}. In order to verify this behavior, we must classify atoms as either bulk or GB atoms. The method by which the GB atoms are separated from the bulk atoms will have an impact on whether bulk atoms exhibit the expected \mbox{$0$ eV} segregation energy. Since a segregation energy spectrum typically only includes GB atoms, the classification scheme will also affect the segregation energy spectra based on the inclusion of atoms near or in the GB that may or may not have segregation energies near \mbox{$0$ eV}.

Bulk atoms are often identified in simulations by adaptive common neighbor analysis (aCNA), as employed in \cite{huber2018machine, mahmood2022atomistic, wagih2019spectrum, wagih2020learning, wagih2023spectrum, cui2023effect, tuchinda2024computed}, which can identify each atom's environment as HCP, BCC, ICO, FCC or other \cite{stukowski2012structure}. Alternatively, the centrosymmetry parameter \mbox{(CSP) \cite{kelchner1998dislocation}} can be used to identify atom environments where the expected centrosymmetry breaks down, such as near a GB in an FCC-type crystal structure. 
There are several less commonly used methods to determine bulk atoms, such as the dislocation extraction algorithm (DXA) \cite{stukowski2010extracting, stukowski2012automated} used in \cite{pei2023atomistic}, the per-site Voronoi volume criterion used in \cite{menon2024atomistic}, and the experimentally determined one-atomic layer region from the GB center \cite{hirokawa1981estimation, guttmann1982thermodynamics}, as employed in \cite{ito2023analysis}.

There are challenges with using the first three of these classification methods because they were designed for purposes other than determining whether an atom belongs to a GB. The aCNA method excels at structure identification, but tolerates distortions of atoms in those structures. The CSP method, since it is continuous, is better suited to differentiate smaller distortions. However, it has no defined cutoff value for discerning when a distortion has changed the structure sufficiently to be classified as something other than FCC, which is left for the user to choose. The other methods suffer from similar challenges, and the experimentally determined one-atomic layer region from the GB center is difficult to compare to simulations. There is no clear way to determine whether an atom definitively belongs to the GB because the transition from bulk to GB can be subtle; elastic strains that cause some deviation from a ``perfect'' bulk structure are present even at large distances.

\begin{figure}[t]
    \centering
    \includegraphics[width=\columnwidth]{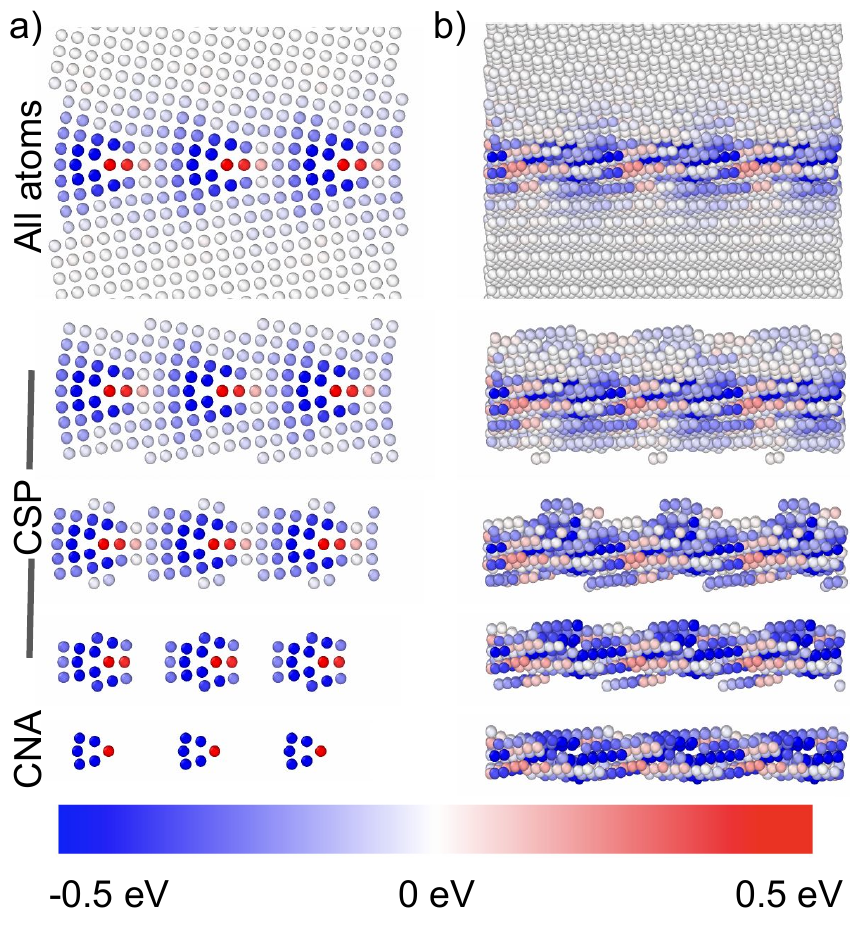} 
    \caption{
    a) A [100] symmetric tilt GB with an array of edge dislocations and 
    b) a high angle, low symmetry GB with ``All atoms'', ``CSP'' bulk atom removal (2nd row: \mbox{CSP $\leq 0.3$}, 3rd row: \mbox{CSP $\leq 0.1$}, 4th row: \mbox{CSP \mbox{$\leq 0.01$}}), and adaptive ``CNA'' bulk atom removal. 
    Red atoms have positive segregation energies, blue have negative, and grey have negligible, according to the colorbar shown. GB images produced using OVITO \cite{stukowski2009visualization}.}
    \label{fig:tableOfGBpics}
\end{figure}

These challenges and the differences between aCNA and CSP classification are illustrated in Figure \ref{fig:tableOfGBpics} for a) a [100] symmetric tilt GB with an array of edge dislocations and for b) a high-angle GB. The atoms are colored according to their segregation energy value (red for anti-segregating, blue for segregating, grey for negligible). The figure depicts the full structure of the GB in the ``All atoms'' row, and the removal of more and more of the surrounding atoms depending on the CSP value used to remove ``bulk'' atoms. The ``CNA'' row shows that the aCNA is aggressive in its removal of ``bulk'' atoms, since its goal is not to identify local distortions in structure but clear changes in crystal structure. The result is that many surrounding atoms with non-negligible segregation energy values are removed by aCNA bulk determination.

Figure \ref{fig:tableOfGBpics} illustrates that there is no definitive approach to atom selection for \mbox{0 eV} bulk GB segregation energies using aCNA or CSP. In this work, we present results for bulk atom selection using both aCNA and CSP methods to compare the impact the selection method has since the aCNA method removes many atoms that could meaningfully contribute to segregation energy spectra.
CSP labeled results use a CSP cutoff of $0.1$, as it includes a reasonable number of surrounding atoms with non-negligible values of the segregation energies while limiting the number of bulk atoms with negligible values of the segregation energy. 

\begin{figure}[t]
    \centering
    \includegraphics[width=\columnwidth]{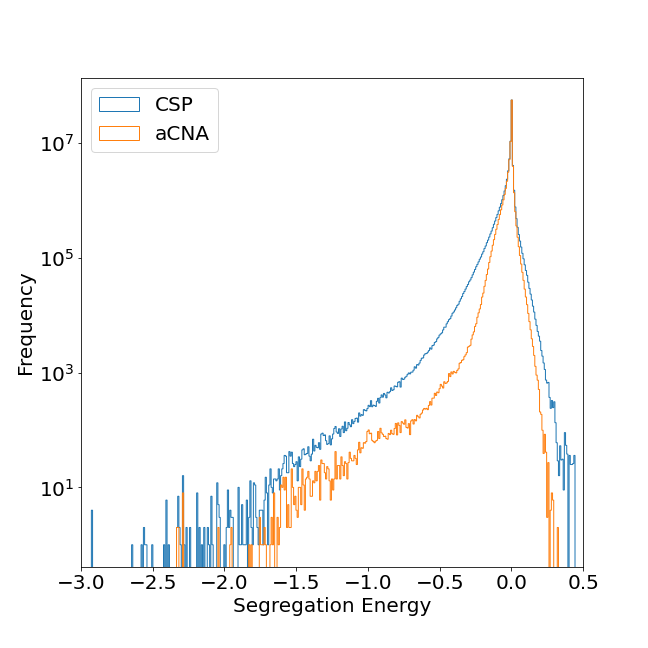}
    \caption{
    Logarithmic scale of the distribution of segregation energies for bulk atoms as determined by CSP $\leq 0.1$ (blue) and aCNA (orange). 
    }
    \label{fig:CSPvsCNAbulk}
\end{figure}

Supplemental Section \ref{sec:CNAvsCSP} contains a discussion comparing bulk atom selection by the aCNA and CSP techniques. The discussion can be summarized in the segregation energy distributions of the bulk atoms by the two classification techniques shown in Figure \ref{fig:CSPvsCNAbulk}. The majority of bulk atoms have segregation energy values close to \mbox{$0$ eV}, though there is a larger than expected variation in the local environments of bulk atoms as determined by both aCNA and CSP. In short, both bulk atom classification schemes classify some atoms as bulk even though they have non-negligible segregation energy values. Given the range of elastic strains near defects, it remains a challenge to find a single defensible method to identify GB and bulk atoms. By contrasting the aCNA and CSP bulk classification schemes in this work, we illustrate the difference between conservative and liberal classification schemes. The determination of a better method for selecting GB atoms is left for the community to resolve.

\begin{figure}[t]
    \centering
    \includegraphics[width=\columnwidth]{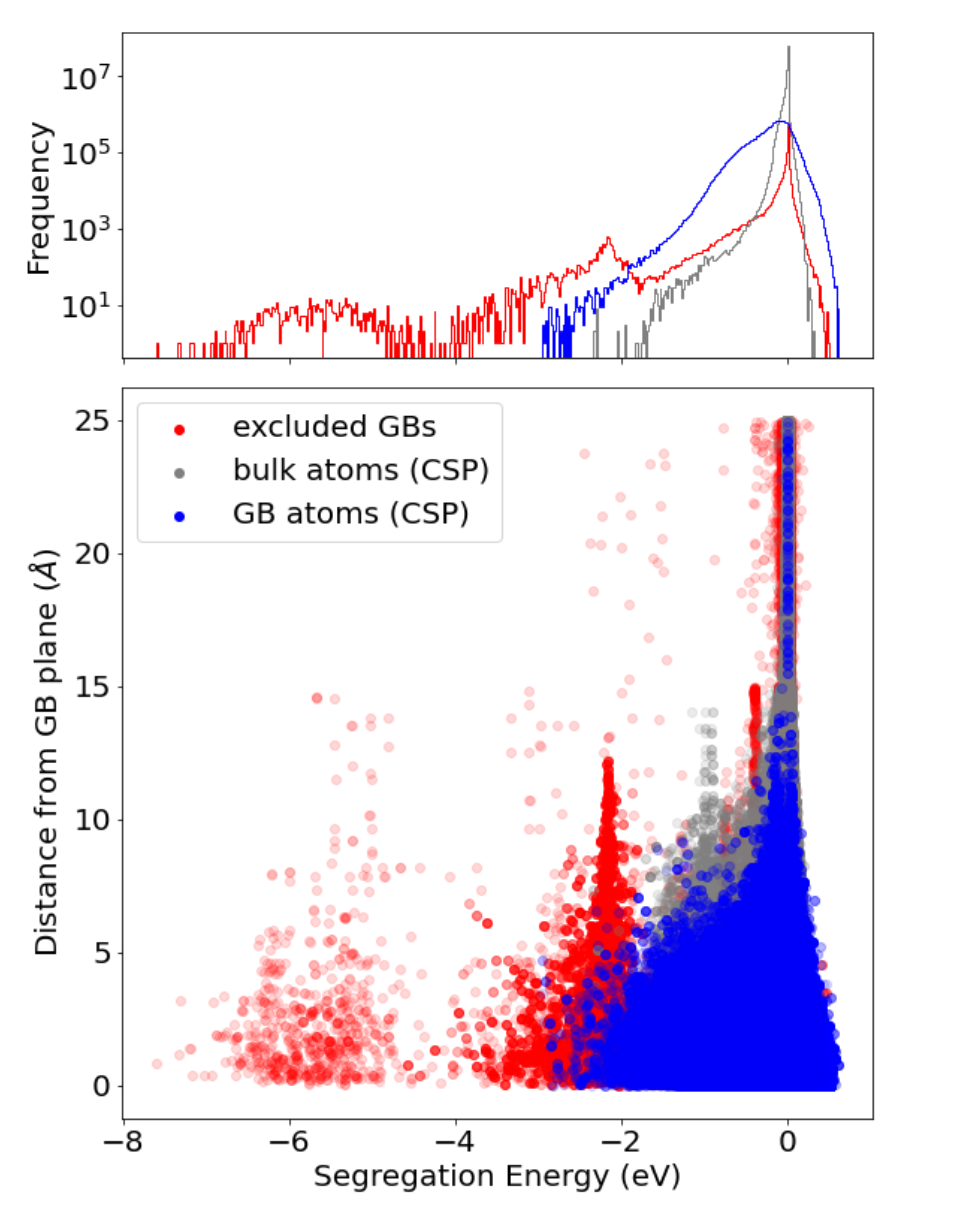} 
    \caption{
    (Top) Histograms of segregation energy for GB atoms, bulk atoms, and atoms in GBs excluded due to invalid calculations. (Bottom) Distance from the GB plane in \AA\ as a function of segregation energy, ($E_{\mathrm{seg}}^{\mathrm{Co}_i}$). In the GB construction simulations, the GB plane is initialized at \mbox{$z=0$ \AA} and is allowed to shift as the simulation cell relaxes, however, this distance is calculated as a distance from \mbox{$z=0$ \AA}. The non-converged and low energy GBs that were removed from analysis as described in Sections \ref{sec:bulkSeg} and  \ref{sec:restructuring} are shown in red, the bulk FCC atoms as determined by a CSP cutoff of $0.1$ are shown in grey, and the non-bulk GB atoms are shown in blue.  The distribution decays to \mbox{$0$ eV} as the distance from the GB increases (with the exception of the excluded GBs, shown in red), as predicted in  \cite{scheiber2015ab, jin2014study}. 
    } 
    \label{fig:segEngVsZ}
\end{figure}

Recalling that theory defines segregation energy of an atom in the bulk as equal to \mbox{$0$ eV}, one would expect segregation energies to converge to \mbox{$0$ eV} as distance from the GB plane increases \cite{scheiber2015ab, jin2014study}. In an initial analysis, it was found that while the majority of data behaved in this way, some did not converge to \mbox{$0$ eV} more than \mbox{15 \AA} away from the GB plane. This is illustrated in a plot of segregation energy vs.\ distance from the GB plane in the scatter plot of Figure \ref{fig:segEngVsZ} for three different populations. These populations make up the \numAtoms atoms and are: the bulk atoms (gray), the GB atoms (blue), and atoms belonging to GBs excluded from the dataset because of errors described here and the following section (red). Four GBs in particular account for the scatter (non-zero segregation energy values) at large distance from the GB, and were therefore excluded from further analysis. These four GBs are listed in Supplemental Table \ref{tab:GBsexcluded} along with GBs excluded for reasons that are discussed in the next section. 

\subsubsection{Challenges due to GB restructuring} \label{sec:restructuring}

It can be seen in the red data points in Figure \ref{fig:segEngVsZ} that there are a small number of very negative segregation energy values, far below what would be expected for this system. When substituting a solute atom for segregation energy calculations, the energy of the new system is calculated after the system has relaxed into the new configuration that accommodates the solute atom. Usually, this relaxation results in almost no change to the atom positions. However, non-negligible changes to the atom positions occur occasionally. The reference energies ($E_\mathrm{Co}^\mathrm{ref}$ and $E_\mathrm{Al}^\mathrm{ref}$ in Equation \ref{eq:segEng}) pertain to the original pre-substitution configuration and not the restructured configuration, therefore the segregation energy values are not valid when significant restructuring occurs. A valid segregation energy value for the restructured configuration would require calculation of new reference energies.

To determine when GB restructuring occurs and the magnitude of such restructuring, we calculated the mean squared displacement (MSD) for all atoms in a GB during the relaxation following solute substitution. Generally, the MSD at GB atom sites was found to be higher than the MSD at FCC bulk atoms (see Supplemental Figure \ref{fig:MSDhist}), but there is considerable overlap between the two distributions. A plot of segregation energy vs.\ MSD is shown in Supplemental Figure \ref{fig:MSDvsSegEng}. 
In this figure it can be seen that i) the most extreme segregation energies occur when the MSD values are higher and i) there are also lots of reasonable segregation energies with relatively large MSD values. Here, we examine examples of these two cases. 

In one case where the segregation energy was very negative, \mbox{$E_{\mathrm{seg}}^{\mathrm{Co}_i} = -6.1$ eV}, and the MSD value was large,  \mbox{$0.076$ \AA$^2$}, considerable restructuring occurred during the post-substitution relaxation. To check the accuracy of this atom's segregation energy value, new reference energy values were determined using the restructured GB for the reference energies. The segregation energy value was re-calculated to be \mbox{$E_{\mathrm{seg}}^{\mathrm{Co}_i} = -0.45$ eV}. Restructuring clearly caused the reference energies to be invalid for the post-substitution GB structure in this case, and MSD was a good determination of this invalidity. 

In another case where the segregation energy was in the normal range, \mbox{$E_{\mathrm{seg}}^{\mathrm{Co}_i} = -0.38$ eV}, but the MSD was still reasonably large, \mbox{$3 \times 10^{-4}$ \AA$^2$}, there was minimal restructuring. The accuracy of this atom's segregation energy value was also checked, and even with new reference energy values, the segregation energy value remained the same at \mbox{$E_{\mathrm{seg}}^{\mathrm{Co}_i} = -0.38$ eV}. In this case, the atomic shuffles had no impact on the segregation energy and the larger than expected MSD values were not indicative of invalid segregation energy values.

Although some high MSD value simulations yield valid segregation energies, we attempted to address the issue of invalid segregation energies caused by restructuring by omitting atoms where MSD values were large. We defined a high-MSD cutoff of \mbox{$10^{-4}$ \AA$^2$}, since it was above this MSD value that the low segregation energies started to diverge (c.f., Supplemental Figure \ref{fig:MSDvsSegEng}). Unfortunately, this approach removed 7 entire GBs and 38\% of the atoms from analysis. Additionally, in some individual GBs, most of the data was lost, as illustrated in Supplemental Figure \ref{fig:tableOfGBpicsMSD}, which shows the effect of removing atoms with high MSD values. 
This approach also removed segregation energy values that were valid, as indicated by the second case examined above. This approach causes severe data loss and could cause misinterpretations of the results. 

Additional analysis showed that most of the extreme (and likely invalid) segregation energy values belonged to a small number of GBs, and these GBs had a high percentage of extreme segregation energy values. In other words, certain GBs were prone to restructuring upon substitution of a solute atom. While not a perfect solution, we removed 18 GBs with segregation energy data less than \mbox{$-3.0$ eV}, which belong to the population of atoms in the excluded GBs shown in red in Figure \ref{fig:segEngVsZ} and listed in Supplemental Table \ref{tab:GBsexcluded}.
Supplemental Figure \ref{fig:NgbsRemoved} shows the segregation energy spectra for other possible segregation energy cutoff values. The chosen cutoff value of \mbox{$-3.0$ eV} removes most of the extreme values while only removing 0.25\% of the GBs simulated. 

While this approach leaves some invalid data in the spectrum due to restructuring in individual simulations, the removal of these 18 GBs seemed the best option, as it only removes 0.25\% of the GBs. We assume that the contributions of any remaining invalid datapoints to the spectrum is minimal, as illustrated in Figure \ref{fig:segEngVsZ}. One way to get around this issue in the future would be to recalculate the reference energy values any time restructuring is detected. Unfortunately, recalculating these reference energies after the fact was impractical for this work.

\subsubsection{Final dataset}

As discussed in Section \ref{sec:bulkSeg}, 4 GBs were removed due to the failure to converge to a zero segregation energy value in the bulk. Another 18 GBs were removed from the dataset because they possessed extremely low segregation energy values, as discussed in Section \ref{sec:restructuring}. There were 10 GBs that were not included due to issues refilling the partially full simulation cells in the dataset. These 32 GBs are listed in Supplemental Table \ref{tab:GBsexcluded}. The remaining data contains \numAtoms atoms with segregation energy values to analyze from \numGBs unique GBs. With bulk atoms removed via CSP or aCNA, this number is reduced to 18 or 11.5 million GB atoms, respectively. Despite the imperfections of these methods as discussed above, it is anticipated that this dataset will provide unique insight into segregation energy trends. Having verified the Homer dataset here, we validate it in the following section.

\begin{figure*}[t]
    \centering
     \includegraphics[width=1.8\columnwidth]{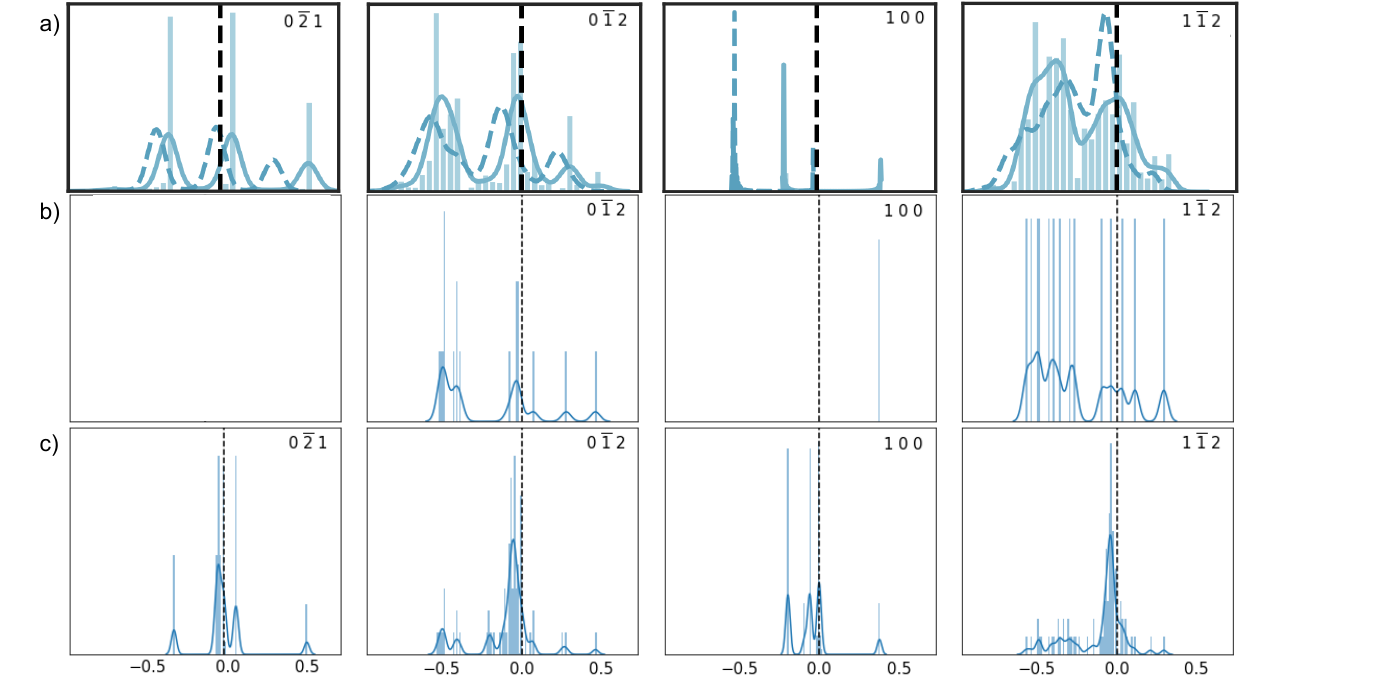}
    \caption{
    Segregation energy distributions of the shared $\Sigma 5$ GBs between the a) Huber dataset from \cite{huber2018machine} and the b/c) Homer dataset. a) has two fit lines: a KDE fit to the data (solid) and a model fit discussed in \cite{huber2018machine} (dashed). b) includes the non-bulk atoms from this work as determined by aCNA (the same method of FCC atom removal as is used in \cite{huber2018machine}). c) includes the non-bulk atoms from this work as determined by CSP. The Huber dataset in a) is reproduced from \cite{huber2018machine} with permission and relabeled according to the Homer dataset labelling conventions \cite{Homer:2022:AlGBdataset}.}
    \label{fig:sigma5}
\end{figure*}

\subsection{Validation} 

In order to validate our work, we compare our results to two computational datasets of Co segregation energies in Al GBs that use the same EAM potential \cite{pun2015interatomic} and similar methods to this work. Huber et al.\ examine a number of $\Sigma5$ GBs in \cite{huber2018machine}, and Wagih et al.\ examine a polycrystal with a variety of different GBs and focus on the overall distribution of segregation energies in \cite{wagih2020learning}. These two datasets are referred to as the Huber dataset and Wagih dataset in the following sections where they are compared with the present results obtained from the Homer dataset.

\subsubsection{Comparison to selected $\Sigma5$ GBs}

Huber et al.\ examine solute segregation in a GB dataset of 38 $\Sigma5 (53.1 ^{\circ} [100])$ GBs \cite{huber2018machine}. The Homer dataset includes 17 GBs of this type, although the two datasets only have four GBs that share all 5 crystallographic degrees of freedom. Figure \ref{fig:sigma5} compares the segregation energy spectra for these four GBs. Figure \ref{fig:sigma5}a is from the Huber dataset with a kernel density estimation (KDE) fit shown with a solid line and a model fit described in \cite{huber2018machine} shown with a dashed line. Figure \ref{fig:sigma5}b is from the present work using the Homer dataset, with segregation energies for non-bulk atoms as determined by aCNA, which is the same method used by Huber et al.\ \cite{huber2018machine}. Figure \ref{fig:sigma5}c is also from the present work using the Homer dataset, but with segregation energies for non-bulk atoms determined by CSP. KDE fits to the distributions in b) and c) are shown with solid lines.

The locations of segregation energy peaks and their relative magnitudes are similar. The $[0 \bar{2} 1]$ GB has the most favorable comparison, though the Huber dataset in a) has a small scattering of infrequent peaks that do not show up in the aCNA Homer dataset in b). In c), the CSP Homer dataset has an additional peak at approximately \mbox{$-0.05$ eV}. The $[0 \bar{1} 2]$ GB has peaks in the same general locations, but slightly different relative magnitudes, again with more scatter in the peaks of the Huber dataset. The CSP Homer dataset again has a large population of atoms with segregation energies near \mbox{$-0.10$ eV} and an additional peak near \mbox{$0.10$ eV} in both b) and c). The $[100]$ GB in our work has a missing peak at approximately \mbox{$-0.55$ eV}, and an additional missing peak near \mbox{$-0.01$ eV} in b) that is present in c). Finally, the $[1 \bar{1} 2]$ GB has more distinct peaks in b) and c) that make up the multimodal distributions around \mbox{$-0.50$ eV}, \mbox{$0.00$ eV}, and \mbox{$0.30$ eV} in a), and a higher relative magnitude in the peak near \mbox{$0.00$ eV}. 

In all of the GBs, c) seems to have more data around \mbox{$-0.05$ eV}, indicating that bulk determination via CSP leaves a larger population of atoms with near-negligible segregation energies that are removed by the aCNA bulk determination method. This is perhaps why aCNA is used in many other works for bulk classification. However, it can be noted that these values are not identically zero, and contribute to the overall relative frequency of the various peaks. This observation is not to advocate for one method over another, but simply to acknowledge a bias introduced by the method of bulk and GB atom selection.

The segregation energy spectra of the two datasets are in general agreement, with appropriate magnitudes and frequency of occurrence. Clearly there is not an exact match, perhaps due to the GBs not having identical structures. Additionally, we cannot guarantee that the aCNA methods have extracted the same population of atoms. For example, the missing peak at approximately \mbox{$-0.55$ eV} in the $[100]$ GB from the Homer dataset when using both aCNA and CSP indicates that the local atomic environment making up that peak in the Huber dataset is not found in the Homer dataset. With no atomic structures from the Huber dataset published, there is no means to verify this conclusion. However, given the fact that we cannot guarantee identical atomic structures, we consider the general agreement of the segregation energy spectra to be sufficient validation of the Homer dataset in comparison with the Huber dataset.

\subsubsection{Comparison to polycrystal spectrum} \label{sec:polycrystalComparison}

While the Homer dataset contains only bicrystal GBs, most materials are polycrystalline, containing a GB network with additional features such as triple junctions and facets. Some recent works have focused on extracting segregation energy distributions from polycrystalline simulations \cite{wagih2019spectrum, wagih2020learning, wagih2023can, Tucker2022, wagih2023spectrum, ito2023analysis}. Here we compare the spectrum of segregation energies obtained from bicrystals of the present work 
with that obtained by Wagih et al.\ from a polycrystal \cite{wagih2020learning}.
The Al-Co segregation energy spectrum from the Wagih dataset is represented by a skew-normal distribution of the form
\begin{equation}
    F(E_{\mathrm{seg}}^{\mathrm{Co}_i}) = \frac{1}{\sqrt{2 \pi} \sigma } \exp \left[ -\frac{\left(E_{\mathrm{seg}}^{\mathrm{Co}_i}-\mu\right)^2}{2 \sigma^2}\right] \mathrm{erfc} \left[ -\frac{\alpha \left(E_{\mathrm{seg}}^{\mathrm{Co}_i} - \mu \right)}{\sqrt{2} \sigma} \right]
    \label{eq:skewnormal}
\end{equation}

\noindent with the fitted location parameter $\mu$, scale parameter $\sigma$, and shape parameter $\alpha$. Note that these values are not the typical mean, standard deviation, and shape parameter of a normal distribution.\footnote{Others have also observed the skew-normal form of the spectrum of segregation energies in GBs, including when bicrystals are used \cite{scheiber2021impact, dosinger2023efficient}.}

\begin{figure}[tb]
    \centering
    \includegraphics[width=\columnwidth]{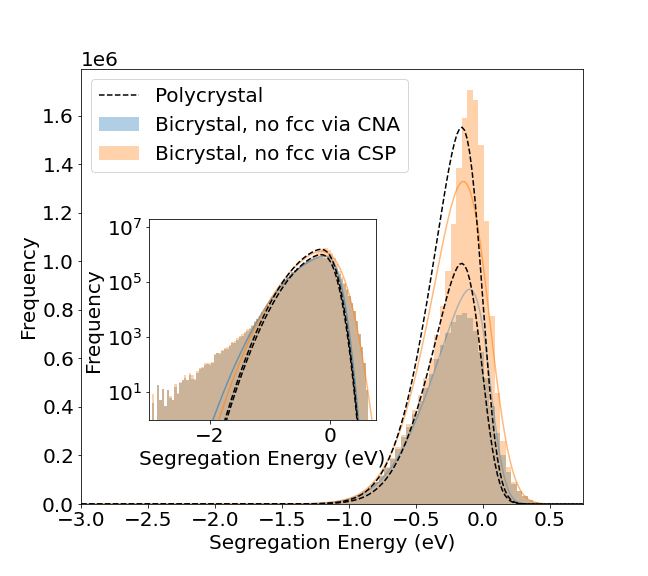}
    \caption{
    The spectrum of segregation energies aggregated from all GBs simulated in this work, with all bulk atoms removed via aCNA (blue) and via CSP (orange), with fit lines for their respective skew-normal distributions in the form of Equation \ref{eq:skewnormal} in their respective colors. The dashed black lines are the polycrystal spectrum from Wagih et al.\ in \cite{wagih2020learning}, scaled to both of this work's distributions. Inset are the same spectra on a logarithmic scale. Statistics for each skew-normal distribution are given in Table \ref{tab:catDataHist}. The spectrum of all atoms (including bulk atoms) is shown in Supplemental Figure \ref{fig:catDataHistwithALL}.} 
    \label{fig:catDataHist}
\end{figure}

The segregation energy spectra for this work are presented in Figure \ref{fig:catDataHist}. The spectra from the Homer dataset are obtained by combining the individual GB segregation energy spectra from all \numGBs GBs. There are two distributions for the two methods by which bulk atoms are removed: aCNA in blue and CSP in orange. The figure includes the Wagih spectrum, scaled to match the two distributions in this work, plotted as dashed black lines, and skew-normal fits in solid lines in corresponding colors to the spectra.  

First, we note that the inclusion of bulk-like atoms using the CSP approach leads to a much larger peak near the origin, though other parts of the histogram of segregation energies are very similar between the aCNA and CSP distributions of the Homer dataset. A comparison of statistical measures of the distributions, in the form of Equation \ref{eq:skewnormal}, are provided in Table \ref{tab:catDataHist}. It can be seen from Table \ref{tab:catDataHist} that the aCNA and CSP distributions of the Homer dataset are in general agreement.

\begin{table}[tb]
\caption{Comparison of skew-normal statistical measures for the distributions shown in Figure \ref{fig:catDataHist}. These parameters\textemdash the fitted location parameter $\mu$, scale parameter $\sigma$, and shape parameter $\alpha$\textemdash are for the skew-normal distribution described in Equation \ref{eq:skewnormal}. Note that these are not the usual mean, standard deviation, or shape parameter of a Gaussian distribution. }
\label{tab:catDataHist}
\begin{tabular}{llll}
\hline
Dataset & $\mu$ & $\sigma$ & $\alpha$ \\ \hline
Polycrystal \cite{wagih2020learning} & -0.0104 & 0.3224 & -3.300 \\
Bicrystals, no FCC via aCNA & 0.0328 & 0.3794 & -2.424 \\ 
Bicrystals, no FCC via CSP & 0.0706 & 0.3502 & -3.365 \\\hline 
\end{tabular}
\end{table}

To compare the datasets to the Wagih dataset, we scaled the magnitude of the skew-normal distribution fitted to the Wagih dataset to match the magnitude of both distributions of the Homer dataset. Note that the Wagih dataset uses aCNA for bulk classification. Despite their quantitative and statistical similarity, there are notable differences in the bicrystal and polycrystal spectra. First, in comparison to the polycrystal spectrum, the bicrystal spectrum that uses aCNA has a slightly higher number of sites with segregation energies just greater than zero and a slightly lower number of sites with segregation energies just less than zero. In the inset with the frequency on a logarithmic scale, it can be seen that there is also significant divergence of both bicrystal spectra from the polycrystal spectrum in the lower tail, but this difference is negligible in the linear frequency scale.

The source of the differences in the two spectra is likely due to the differences in the datasets. First, the polycrystal dataset has 16 distinct grains, 72 GBs, and ${\sim}10^5$ non-bulk GB atoms \cite{wagih2020learning}, in comparison to the $2\times\numGBs$ distinct grains, \numGBs GBs, and ${\sim}10^6$ non-bulk GB atoms for the bicrystals in the present work. In addition, the polycrystalline simulation has atomic environments found in GB triple junctions and quadruple nodes that may not appear in bicrystal simulations, or if those environments show up in bicrystal GB simulations, they may appear with a different frequency than they do in a polycrystalline structure. 

It has been shown that small populations of bicrystal GBs fail to produce the same segregation energy distributions as those of polycrystalline materials \cite{wagih2023can, Tucker2022}, but that as the population diversity increases with GBs of lower symmetry, there is better coverage of the atomic environment space \cite{wagih2023can}. The present dataset of \numGBs bicrystal GBs is comprised of mostly low symmetry GBs; only 89 of the \numGBs grain boundaries, or 1.2\%, have low CSL values (i.e., $\Sigma \leq10$), suggesting that issues related to diversity in the dataset are minimized by the large variety of GBs in the Homer dataset.

Conversely, the polycrystal simulation with its 72 GBs may not provide adequate sampling of the variation in structure across the 5D space, or be large enough to be considered a representative volume element (RVE) such that it is truly representative of atomic environments in a polycrystal. The Mackenzie Distribution, which represents the distribution of disorientation angles for a polycrystalline sample with random cubic crystal orientations \cite{Mackenzie.1957.Biometrika, Mackenzie.1958.Biometrika}, can be used as a justification for the selection of an RVE \cite{HUSSEIN2021116799}. As shown in Figure 3 of \cite{Homer:2022}, the Homer bicrystal dataset gives a reasonable approximation of the Mackenzie Distribution. The Wagih polycrystal dataset does not claim to follow the Mackenzie Distribution of disorientations, nor do they consider their simulation to be an RVE. Wagih et al.\ do however assert that their simulations are similar enough to randomly oriented grains to represent the local atomic environments present in the polycrystalline GB space, and that the segregation energy spectrum obtained is universal to any segregation energy spectrum obtained from a polycrystal \cite{wagih2019spectrum, wagih2023can}.

At this point, it is unclear if the differences between the Wagih and aCNA bicrystal distributions are significant. The degree to which either of the methods incorporates aspects of the distribution of GBs that are critical to a proper representation of segregation energies in a diverse polycrystal is also unclear. The discussion leads to several unanswered questions: 
\begin{enumerate}
    \item How many GB types would be needed to establish an RVE for segregation energies? (i.e., is the polycrystal sample from Wagih et al.\ large enough? Does the present work have enough and sufficiently diverse bicrystals to represent a polycrystal?)
    \item How would a change in GB texture of the polycrystal change the GB sampling and thus the segregation energy spectrum? (i.e., does the texture used in the work of Wagih et al.\ bias the sampled spectrum significantly? Is it appropriate to make these distributions from a random sampling?) 
    \item To what degree do the local atomic environments of triple junctions and quadruple nodes affect the sampled segregation energy spectrum? (i.e., the volume fraction of such atomic environments will be different in nanometer-sized grains as compared with micron-sized grains, and may be entirely absent from the bicrystal dataset of this work.)
\end{enumerate} \label{tab:questions}
\noindent We leave these questions to the community to address. Nevertheless, the similarity in the two spectra is seen as a positive validation of the methods and use of the Homer GB dataset in this work.

\section{Results \& Discussion} \label{sec:results}

Having provided some context and discussion of the results in the Verification and Validation section, we begin our analysis here by taking several different views of the segregation energy spectra produced in this work. Note that most of the results presented in this section are produced using CSP to determine bulk atoms, but similar results would be obtained using aCNA, as will be described in Section \ref{sec:5Dtrends}. 

Examining the spectra of segregation energies across the 5D space is challenging because at each point in the space we obtain a spectrum of segregation energies. To provide insight into the dataset, we employ several different tactics: i) statistical measures of the spectra, i) classification of the segregation energy spectra into fractions of atoms segregating, anti-segregating and negligibly segregating, and i) calculation of a GB solute concentration for each GB ($c_\mathrm{GB}$ defined in Equation \ref{eq:conc}) based on Coghlan and White's model \cite{white1977spectrum}. Each of these provide different insight into trends in the large set of GBs.

\subsection{Statistical measures of the spectra} \label{sec:maxMeanMin}

In this section, to examine the segregation energy spectra across all of the dataset, we present the mean, maximum, and minimum values of the multimodal distributions of each GB's segregation energy spectrum. As an example, the maximum, mean, and minimum values for the $\Sigma 5$ GB spectra shown in Figure \ref{fig:sigma5}c are given in Table \ref{tab:sigma5stats}.

\begin{table}
\caption{Statistics for the $\Sigma5$ GBs shown in Figure \ref{fig:sigma5}. Note that the the max, mean, and min values in the table refer to the maximum, mean, and minimum values from each GB's $E_{\mathrm{seg}}^{\mathrm{Co}_i}$ spectrum, given in eV. $c_\mathrm{GB}$ is given in at\%.}
\centering
\begin{tabular}{cccccccc} \hline
   GB & max & mean & min & $f_\mathrm{seg}$ & $f_\mathrm{negl}$ & $f_\mathrm{anti}$ & $c_\mathrm{GB}$ \\
    \hline   
   $0\overline{2}1$  & 0.52 & -0.01 & -0.31 & 0.29 & 0.29 & 0.43 & 28.9 \\ 
   $0\overline{1}2$ & 0.46 & -0.14 & -0.53 & 0.47 & 0.40 & 0.14 & 37.7 \\ 
   $100$ &0.34 & -0.05 & -0.20& 0.44 & 0.44 & 0.11 & 35.5\\ 
   $1\overline{1}2$  & 0.30 & -0.18 & -0.57 & 0.58 & 0.24 & 0.18 & 59.7 \\ 
   \hline
\end{tabular}
\label{tab:sigma5stats}
\end{table}

Although this represents a significant reduction of information, some observations can be still made by comparing these values against one-dimensional parameterizations of the GB dataset. For example, these values are plotted as a function of GB interface energy, $\gamma$, in Figure \ref{fig:segVsGamma} and disorientation angle in Supplemental Figure \ref{fig:segVsDis}. 
Generally, the mean segregation energy of a GB becomes more negative as GB energy increases, as shown by the green datapoints and trendline in Figure \ref{fig:segVsGamma}. Note that the [111] symmetric twist GBs (aside from the lowest energy perfect twin GB) have a slight increase of the mean segregation energy as GB energy increases, contrary to the general trend. We will further analyze this subset of GBs in Section \ref{sec:soluteConc}. The average range of segregation energies also increases as a function of GB energy, as indicated by the linear fits to the three populations shown in black. This is probably due to greater deviation from the bulk FCC structure in the higher energy GBs, which likely leads to a greater distribution of segregation energies in the solute atom sites. Finally, it can be seen in Figure \ref{fig:segVsGamma} that the lower the mean segregation energy for a GB, the higher the probability for segregation in that GB. 

\begin{figure}[t]
    \centering
    \includegraphics[width=\columnwidth]{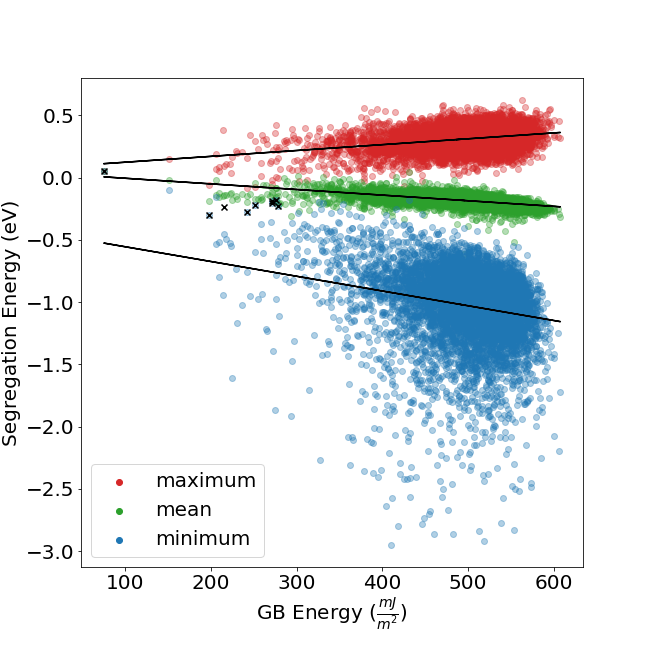}
    \caption{
    Maximum (red), mean (green), and minimum (blue) values of the segregation energy, $E_{\mathrm{seg}}^{\mathrm{Co}_i}$, spectrum for each GB vs.\ their GB interface energy, $\gamma$. Linear fits for each are shown in black. The mean segregation energy for the [111] symmetric twist GBs are shown with black ``x'' markings. 
    The range of segregation energies increases with GB energy, and the mean segregation energy has an inverse relationship with increasing GB energy.} 
    \label{fig:segVsGamma}
\end{figure}

GB energy, $\gamma$, has been shown to have a Read-Shockley relationship with disorientation angle \cite{Read:1950um, Wolf:1989ci}. Despite the relationship observed between GB energy and mean segregation energy, there is no observable transitive relationship between mean segregation energy and disorientation angle (see Supplemental Figure \ref{fig:segVsDis}). It is interesting that the two separate correlations do not result in the correlation between segregation energy and disorientation angle. While the correlation of segregation energy with GB energy provides some insight to global trends in segregation energy, we continue with more detailed analyses of the dataset.

\subsection{Classification of the spectra} \label{sec:fractions}


To examine the segregation energy spectra in this section, we employ the segregation energy classifications described in Section \ref{sec:methods-analysis}. This approach classifies each atom based on its segregation energy value into one of three categories: i) segregating (\mbox{$E_{\mathrm{seg}} ^{\mathrm{Co}_i} < -0.0875$ eV}), ii) negligibly segregating (\mbox{$-0.0875$ eV $\leq E_{\mathrm{seg}} ^{\mathrm{Co}_i} < +0.018$ eV}), or iii) anti-segregating (\mbox{$E_{\mathrm{seg}} ^{\mathrm{Co}_i} \geq +0.018$ eV}). Using these classifications, we compute the fraction of atoms in each GB that fall into each category, i) $f_\mathrm{seg}$, ii) $f_\mathrm{negl}$, and iii) $f_\mathrm{anti}$.\ For each GB, these fractions add to 1 (i.e., $f_\mathrm{seg} + f_\mathrm{negl} + f_\mathrm{anti} = 1$). As an example, the $f_\mathrm{seg}$, $f_\mathrm{negl}$, and $f_\mathrm{anti}$ values for the $\Sigma 5$ GB spectra shown in Figure \ref{fig:sigma5}c are given in Table \ref{tab:sigma5stats}.

\begin{figure}[t]
    \centering
    \includegraphics[width=\columnwidth]{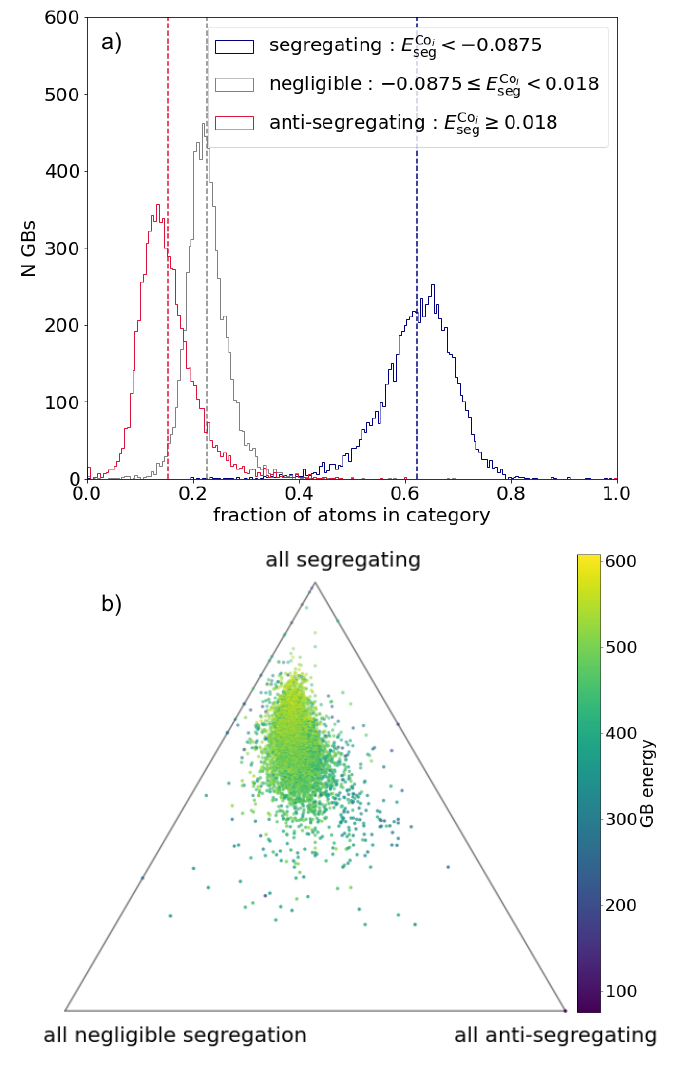}
    \caption{
    a) Histograms of $f_\mathrm{seg}$ (blue), $f_\mathrm{anti}$ (red), $f_\mathrm{negl}$ (grey) for GB atoms as determined by CSP. The mean of each histogram is marked with a dashed vertical line of a corresponding color.
    b) Ternary scatter plot of the same data, colored by GB energy, $\gamma$. Since the sum of these values is 1 (i.e., $f_\mathrm{seg} + f_\mathrm{anti} + f_\mathrm{negl} = 1$), they lie on a 2D plane in 3D space.} 
    \label{fig:histOfFracSeg}
\end{figure}

The distributions of these fractions for the \numGBs GBs are shown in Figure \ref{fig:histOfFracSeg}a; these are the distributions for non-bulk atoms identified using the CSP scheme. The mean value for these categories over all GBs are $\overline{f_\mathrm{seg}} = 0.62$, $\overline{f_\mathrm{anti}} = 0.15$, and $\overline{f_\mathrm{negl}} = 0.23$, as shown by the dotted vertical lines in Figure \ref{fig:histOfFracSeg}a. The global fractions in the aggregated spectrum are $f_\mathrm{seg} = 0.63$, $f_\mathrm{anti} = 0.15$, and $f_\mathrm{negl} = 0.22$. 

In Figure \ref{fig:histOfFracSeg}b, this same data is plotted on a ternary plot, that shows the 2D plane that the 3D points all lie on, since they all add to 1 (i.e., $f_\mathrm{seg} + f_\mathrm{negl} + f_\mathrm{anti} = 1$). The data points in the plot are colored by GB energy, $\gamma$. It can be seen in Figure \ref{fig:histOfFracSeg}b that increased GB energy corresponds to an increased fraction of segregating atoms, $f_\mathrm{seg}$. This supports the trend for more negative segregation energies with higher GB energy observed in Figure \ref{fig:segVsGamma}. To see the same plots for aCNA, see Supplemental Figure \ref{fig:ternaryPlotCNA}. 

As expected from the histograms, the general distribution of points in Figure \ref{fig:histOfFracSeg}b is located closest to the ``all segregating'' corner of the triangle. There are several notable outliers. The [111] perfect twin GB is located at ``all anti-segregating'', and its closest neighbor in the ternary plot has the same misorientation of $60^{\circ}$ about the [111] axis. [100], [110] and [111] disorientation axis GBs make up 50\% of the GBs along and near the top right edge of the triangle with $f_\mathrm{negl} \leq .1$. 
34\% of the GBs with $f_\mathrm{anti} \leq .075$ belong to the [111] disorientation axis.\footnote{To see the ternary plot with the [111] disorientation axis GBs highlighted, see Supplemental Figure \ref{fig:ternaryPlot111}.} Thus, many of the outliers in the ternary plot belong to high symmetry disorientation axes. 
Additionally, it can be seen that there are a number of GBs along the left side of the plot which have no anti-segregating atoms, though this is perhaps unsurprising given the proximity of the distribution to this edge and the tendency to segregate in this system.  

In short, most GBs have a tendency to segregate, evidenced by the $\overline{f_\mathrm{seg}}$ value of $0.62$. Additionally, high energy GBs have a tendency to segregate more than lower energy GBs, which is supported by the results shown in Figure \ref{fig:segVsGamma}. However, while these trends are interesting, this analysis is still insufficient to predict segregation behavior generally.

\subsection{Solute concentration at GBs} \label{sec:soluteConc}

One of the challenges with the representations of the segregation energy spectra analyzed in the preceding sections is that they still have multiple values for each GB, which makes it hard to analyze trends across the 5D space. As such, we have included the calculation of the grain boundary solute concentration, $c_\mathrm{GB}$, from Equation \ref{eq:conc}. This singular value for each GB allows the display of some trends more clearly in subsets of the 5D space that still take into account attributes of the full spectrum of GB energies, as the value is computed from a sum over the whole spectrum of GB atoms in each GB. Note that most of the $c_\mathrm{GB}$ values presented in this section are well above the dilute limit; as such, many of the assumptions made in the calculation of the concentration values (e.g., neglecting solute-solute interactions) are invalid. However, we assume that the general trends observed in $c_\mathrm{GB}$ are still valid. 

\begin{figure}[h!]
    \centering
    \includegraphics[width=\columnwidth]{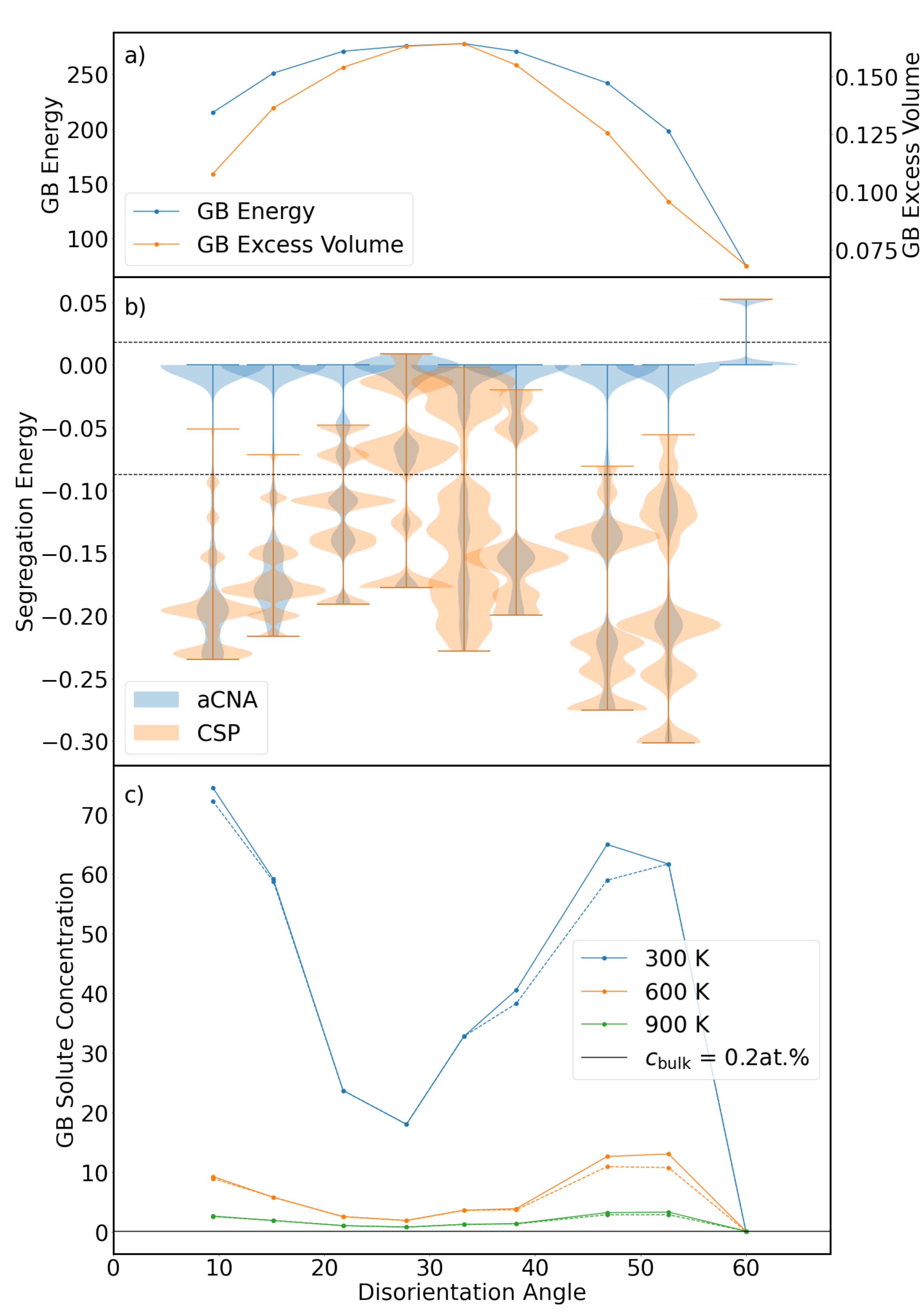}
    \caption{
    a) Plot of GB energy ($\mathrm{mJ}/\mathrm{m}^2$) and GB excess volume ($\AA^3/\AA^2$) as functions of disorientation angle ($^\circ$) for the [111] symmetric twist GBs.
    b) Plot of the segregation energy (eV) spectra for GB atoms as determined by CSP (orange) and aCNA (blue) in the [111] symmetric twist GBs, against their disorientation angles. Aside from the perfect twin GB at $60^{\circ}$ twist, their segregation energies are almost always negative. The ``negligible segregation" (\mbox{$-0.0875$ eV $\leq E_{\mathrm{seg}} ^{\mathrm{Co}_i} < +0.018$ eV}) limits are shown with dotted black lines.
    c) Concentration of solute at the GB, $c_{\mathrm{GB}}$ (at\%), as a function of temperature for the [111] symmetric twist GBs, for CSP (solid) and aCNA (dotted). This is calculated using Equation \ref{eq:conc}. The bulk concentration, \mbox{$c_{\mathrm{bulk}}$=0.2at\%}, is shown with a solid black line. } 
    \label{fig:SymTwistGBs}
\end{figure}

As with the other sections, the $c_\mathrm{GB}$ values for the $\Sigma 5$ GB spectra shown in Figure \ref{fig:sigma5}c are given in Table \ref{tab:sigma5stats}. However, because $c_\mathrm{GB}$ takes into account the full spectrum, we use a set of GBs that have a unique trajectory through the 5D space of GB crystallographic character to illustrate the $c_\mathrm{GB}$ values. This set of GBs is the low GB energy [111] symmetric twist GBs noted in \cite{Homer:2022} and mentioned in Section \ref{sec:maxMeanMin}, used because they have a large range of disorientation angles, and because they show interesting segregation energy results. Note that the disorientation angles related to these particular boundaries also define the twist angles about the [111] axis. 

The [111] symmetric twist GBs exhibit an energy trend of an inverted parabola with low GB energy values at low and high disorientation angles, illustrated in blue in Figure \ref{fig:SymTwistGBs}a. A corresponding trend for GB excess volume is also shown in orange in Figure \ref{fig:SymTwistGBs}a. Figure \ref{fig:SymTwistGBs}b plots the segregation energy spectra for the GB atoms as determined by aCNA (blue) and CSP (orange) for each of these GBs, against their disorientation angle. Note the data loss from using aCNA (blue) over CSP (orange), which was discussed in Section \ref{sec:bulkSeg}. Although a significant portion of each GB's spectra falls in the ``negligible segregation'' category shown by horizontal dotted black lines in Figure \ref{fig:SymTwistGBs}b, a majority do not, and almost all of the non-negligible segregation energies are negative, implying that a Co atom added to an Al [111] symmetric twist GB will prefer to segregate to the GB. 
With the exception of the perfect twin, this segregation data shown in Figure \ref{fig:SymTwistGBs}b generally runs counter to the general trend that was observed in Figure \ref{fig:segVsGamma} for lower energy GBs to have less segregation than higher energy GBs. This is evidenced by the more negative segregation energy spectra in Figure \ref{fig:SymTwistGBs}b for the corresponding lower GB energies in Figure \ref{fig:SymTwistGBs}a.

In Figure \ref{fig:SymTwistGBs}c, we show the $c_\mathrm{GB}$ values as determined by CSP (solid) and aCNA (dotted) for the [111] symmetric twist GBs as a function of disorientation angle at 3 different temperatures. In addition to the low energy GBs in this dataset having more negative segregation energies, they also have higher solute concentrations, excepting the perfect twin at $60^{\circ}$. This is seen most clearly in the $c_\mathrm{GB}$ line for 300 K (orange) in Figure \ref{fig:SymTwistGBs}c. This correlation is expected from the inverse relationship between segregation energy, $E_{\mathrm{seg}} ^{\mathrm{Co}_i}$, and solute concentration, $c_\mathrm{GB}$ in Equation \ref{eq:conc}, and can be seen clearly by the paraboloid quality of the segregation energy spectra across disorientation angles in Figure \ref{fig:SymTwistGBs}b with a matching inverse paraboloid in Figure \ref{fig:SymTwistGBs}c.

As demonstrated in Figure \ref{fig:SymTwistGBs}c and expected from Equation \ref{eq:conc}, the concentration of solute at the GB, $c_\mathrm{GB}$, is temperature dependent. Note that segregation energies were calculated at $0$ K, but that Equation \ref{eq:conc} expects this to be the case. We expect that as the temperature increases, less favorable atom sites in the bulk of the material are occupied with increasing probability.
This causes the concentration to approach the bulk value at elevated temperatures, which is shown clearly by the decreasing concentration values at higher temperatures in Figure \ref{fig:SymTwistGBs}c. Thus the concentrations calculated here match the theory presented in Section \ref{sec:methods-analysis}.

The perfect twin GB at a $60^{\circ}$ twist angle about the [111] disorientation axis is an exception to most of the trends in the [111] symmetric twist GBs discussed here. For example, it has all positive segregation energies, shown in Figure \ref{fig:SymTwistGBs}b. This is unsurprising given the structure of the twin boundary; its density and structure provide no easy sites for segregation as compared with the bulk (evidenced by its low excess volume), leading to positive segregation energy values in all cases. The highly symmetric structure also results in a very low GB energy, shown in Figure \ref{fig:SymTwistGBs}a. The perfect twin has lower $c_\mathrm{GB}$ values than the rest of the [111] symmetric twist GBs at every temperature, as shown in Figure \ref{fig:SymTwistGBs}c, as a result of its entirely positive segregation energy spectrum. It is also the only GB in the Homer dataset with a $c_\mathrm{GB}$ value below $c_\mathrm{bulk}=0.2$at\%, having a value of $c_\mathrm{GB}=0.03$at\% for T=300K.

It can be seen that aCNA and CSP give similar results at \mbox{300 K} (compare the 300 K solid and dotted lines in Figure \ref{fig:SymTwistGBs}c). This was found to be the case for the entire dataset; the correlation between $c_\mathrm{GB}$ computed from CSP vs.\ aCNA GB atoms is plotted in Supplemental Figure \ref{fig:concCSPvsCNA}. Using CSP nearly always results a slightly lower $c_\mathrm{GB}$ value since it includes more near-bulk atoms than aCNA, but the positive correlation means that all trends should remain the same.

\begin{figure*}[h]
    \centering
    \includegraphics[width=\textwidth]{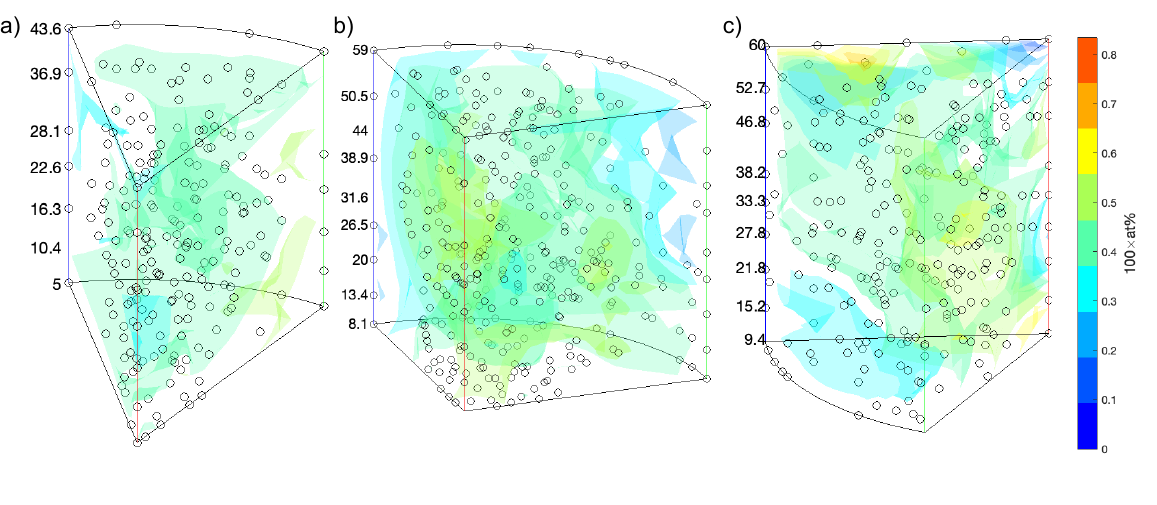}
    \caption{
    Disorientation angle vs.\ boundary plane volumetric plots of concentration in the GB, $c_\mathrm{GB}$, with isosurfaces colored by different $c_\mathrm{GB}$ values for 3 different disorientation axes: a) $[100]$ b) $[110]$ c) $[111]$. These plots are equivalent to stacking boundary plane fundamental zone plots with the same disorientation axes. The [111] symmetric twist GBs are located along the vertical red line in plot c), with the perfect twin located at the top, corresponding to the darkest blue contour, and the lowest $c_\mathrm{GB}$. }
    \label{fig:cakePlots}
\end{figure*}

Having examined the [111] symmetric twist GBs, we now turn our attention to the $c_{\mathrm{GB}}$ values for all \numGBs GBs over the 5D space. 82.1\% of non-FCC GB atom sites observed in this work, as determined by CSP, have negative segregation energies, which corresponds to segregation being favorable. Therefore, a majority of GB sites will accommodate a Co atom in the dilute limit. 
This implies that the concentration of Co atoms in the GB, $c_{\mathrm{GB}}$, will be higher than in bulk, and is evidenced by the mean concentration of all GBs in the Homer dataset, which is $\overline{c_\mathrm{GB}} = 46.7 \pm 7.2$ at\%. 
Supplemental Figure \ref{fig:concVsGamma} shows $c_{\mathrm{GB}}$ vs.\ GB energy for all GBs in the Homer dataset.

\begin{table}
\centering
\caption{Disorientation axes with the highest mean $c_\mathrm{GB}$ values, and the number of GBs that belong to that axis.} \label{tab:highconc}
\begin{tabular}{ccc}
    \hline
    Disorientation Axis & $\overline{c_\mathrm{GB}}$ for this axis & \# of GBs \\\hline
    $[443]$ & $52.7 \pm 4.6$ at\% & 86 \\
    $[751]$ & $51.4 \pm 5.5$ at\% & 108\\
    $[654]$ & $51.3 \pm 7.7$ at\% & 20\\\hline
\end{tabular}
\end{table}

\begin{table}
\centering
\caption{Disorientation axes with the lowest mean $c_\mathrm{GB}$ values, and the number of GBs that belong to that axis.} \label{tab:lowconc}
\begin{tabular}{ccc}
    \hline
    Disorientation Axis & $\overline{c_\mathrm{GB}}$ for this axis & \# of GBs\\\hline
    $[100]$ & $40.6 \pm 6.7$ at\% & 217 \\
    $[554]$ & $41.1 \pm 8.5$ at\% & 32\\
    $[110]$ & $42.6 \pm 10.4$ at\% & 352\\
    $[111]$ & $43.0 \pm 11.6$ at\% & 253\\\hline
\end{tabular}
\end{table}

The disorientation axes with the highest and lowest mean $c_\mathrm{GB}$ values are given in Tables \ref{tab:highconc} and \ref{tab:lowconc}, respectively.
It is worth noting that the low $c_\mathrm{GB}$ disorientation axes in Table \ref{tab:lowconc} have high symmetry, with the exception of the [554] axis. 
However, it is also worth noting that the mean $c_\mathrm{GB}$ for the entire dataset lies within one standard deviation of each of both the low and high concentration disorientation axes' mean values given in Tables \ref{tab:highconc} \& \ref{tab:lowconc}, aside from the [443] axis. Thus, there is considerable overlap in the distributions. Additionally some of the axes listed here may not be statistically significant enough to be considered outliers, based on their small populations (e.g., the [654] GB containing only 20 GBs). So, while they may have more extreme $c_\mathrm{GB}$ values in general and also contain some of the GBs with outlying ternary plot locations mentioned in Section \ref{sec:fractions}, 
the actual range of $c_\mathrm{GB}$ values in all cases is not large and it is difficult to find meaningful trends among the averaged values of the disorientation axes.

The mean concentrations of all of the Homer dataset GBs at each CSL are plotted in Rodriguez space  in Supplemental Figure \ref{fig:concRFspace}.\footnote{Rodriguez space is also known as Rodriguez-Frank space and is a fundamental zone where the CSL values of cubic-cubic disorientation GBs are defined. It is a 3D parameterization of disorientation, and as such is often used to aid in visualization of 5D datasets of GBs.} Consistent with Table \ref{tab:lowconc}, the [100], [110], and [111] disorientation axis GBs have the lowest $\overline{c_\mathrm{GB}}$ values per disorientation axis, although the [111] axis has a larger deviation. The smooth variation of $c_\mathrm{GB}$ in Rodriguez space is an indicator that there may be broader global trends in $c_\mathrm{GB}$, but because these represent averages of dozens of $c_\mathrm{GB}$ values for the different boundary planes, any functional would be complex. 

The effect of boundary plane is illustrated in Figure \ref{fig:cakePlots}, which shows three small subsections of the 5D GB space\textemdash the a)$[100]$, b)$[110]$, and c)$[111]$ disorientation axis GBs. These plots are volumetric plots of concentration, $c_\mathrm{GB}$, where the z-axis defines the disorientation angle and the x- and y-axes define a stereographic projection of the boundary plane normal in boundary plane fundamental zones. The vertices of the plots define high symmetry boundaries; the two vertices that terminate each arc are symmetric tilt boundaries about the disorientation axis and the other vertex defines symmetric twist boundaries about the disorientation axis. The [111] symmetric twist boundaries examined in Figure \ref{fig:SymTwistGBs} correspond to the points along the red line in Figure \ref{fig:cakePlots}c.

All three plots in Figure \ref{fig:cakePlots} show smooth but unpredictable variation of the $c_\mathrm{GB}$ from areas with high $c_\mathrm{GB}$ to other areas of low $c_\mathrm{GB}$. We refer to this as a complex or rugged landscape \cite{palmer2018optimization}. This means that while $c_\mathrm{GB}$ varies smoothly, there are many irregular local extrema and a lack of symmetry or global trends. Of the three axes shown in Figure \ref{fig:cakePlots}, the c) $[111]$ disorientation axis exhibits the largest range of $c_\mathrm{GB}$. This is perhaps unsurprising because it has the low energy twin GB at the top of the red line, and Table \ref{tab:lowconc} shows that it has a large standard deviation of 11.6 at\%. 
Additionally, as noted above, GBs from the three axes in Figure \ref{fig:cakePlots} make up many of the outliers in the ternary plot in Figure \ref{fig:histOfFracSeg}b, which may be related to their low mean $c_\mathrm{GB}$ values (noted in Table \ref{tab:lowconc}) as compared to the global mean concentration of $\overline{c_\mathrm{GB}} = 46.7 $at\%. However, as stated earlier, not every GB from these axes will be an outlier. 

While there are not broad global trends that we can extract from these few subspaces analyzed, these plots illustrate the effects of disorientation and boundary plane on changes to segregation energy spectra in grain boundaries. The segregation energy spectra or $c_\mathrm{GB}$ values computed in this work could be used to develop a model for segregation across the 5D space (e.g., using an expansion \cite{mason2019basis} or an interpolation function \cite{Baird:2022:GBenergyBarycentricInterpolation}). Such a model could subsequently be used to examine the effects of texture or estimate segregation for a GB of arbitrary character.

\subsection{Overall trends in dataset} \label{sec:5Dtrends}

The broad effect of GB crystallographic character on GB segregation trends has to this point been unknown and was recently listed as a future perspective worth considering \cite{hu2024computational}. The plots of $c_\mathrm{GB}$ in Figure \ref{fig:cakePlots} and Supplemental Figure \ref{fig:concSymmetric} show that segregation varies smoothly throughout the 5D crystallographic space. Unfortunately, the landscape is rugged and does not indicate any obvious trends of segregation as a function of 5D crystallographic character. However, $c_\mathrm{GB}$ and the segregation energy spectrum do appear to correlate with GB energy and GB excess volume at the local scale, as illustrated by the [111] symmetric twist GBs in Figure \ref{fig:SymTwistGBs}. In addition, segregation energies appear to be correlated with GB energy across the whole dataset as illustrated in Figure \ref{fig:segVsGamma}. This is supported by observations from Huber et al.\ in their exploration of $\Sigma 5$ GBs, who found that segregation energy per site depended on excess volume and coordination number at the site.  

\begin{figure}[ht]
    \centering
    \includegraphics[width=\columnwidth]{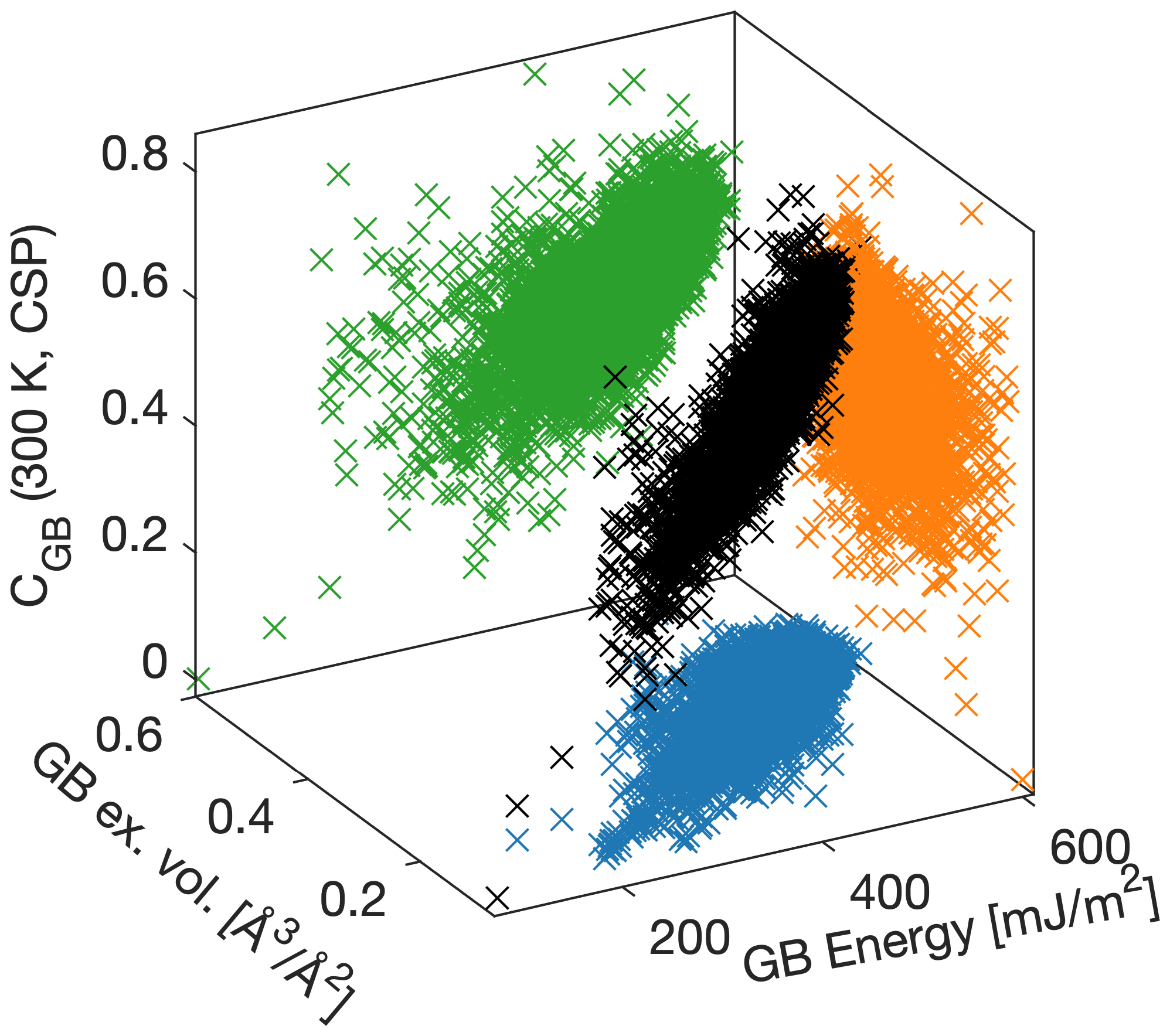}
    \caption{Concentration of solute at the GB, $c_\mathrm{GB}$, as a function of GB excess volume and GB energy. The following are projected onto their corresponding planes: $c_\mathrm{GB}$ vs.\ GB energy (green), $c_\mathrm{GB}$ vs.\ GB excess volume (orange), and GB excess volume vs.\ GB energy (blue). }
    \label{fig:concVsEandEV}
\end{figure}

\begin{table}
\caption{Estimate, standard error (SE), and $p$-value of variables of the following linear model used to predict $c_\mathrm{GB}$: 
$c_\mathrm{GB} = x_0 + x_1 \cdot \gamma + x_2 \cdot V_\mathrm{exc} + x_3 \cdot \theta_\mathrm{dis}$.}
\centering
\begin{tabular}{cccc} \hline
    Variable & Estimate    &   SE  &   $p-$value  \\
    \hline
    $x_0$   &   0.0708   &   0.00496   &   1.37$\times 10^{-45}$ \\
    $x_1$   &   0.00175 &   1.54$\times 10^{-5}$    &   0   \\
    $x_2$   &  -1.42    &   0.0163      &   0   \\
    $x_3$   &   5.07$\times 10^{-5}$    &   4.10$\times 10^{-5}$    &   0.21698 \\\hline
\end{tabular}
\label{tab:linearmodel}
\end{table}

Given these apparent correlations, we created a simple linear model to predict $c_\mathrm{GB}$. We include disorientation angle to search for possible crystallographic dependence as well as two variables known to correlate with segregation, GB energy and GB excess volume. A plot of these variables, excluding disorientation angle, is shown in Figure \ref{fig:concVsEandEV}. The linear model is given in Table \ref{tab:linearmodel}. The model has an $R^2$ value of 0.642 and a root mean squared error (RMSE) value of 0.043 eV. In contrast, a linear model of $c_\mathrm{GB}$ as a function of GB energy alone, shown in Supplemental Figure \ref{fig:concVsGamma}, has an $R^2$ value of 0.271. A plot of $c_\mathrm{GB}$ vs.\ disorientation angle is shown in Supplemental Figure \ref{fig:concVsDis} to confirm the lack of correlation with $\theta_\mathrm{dis}$ found in the model; the p-value on the coefficient is $0.217$. This result contrasts the strong misorientation dependence observed in a small sampling of GBs for a Pt–Au sample \cite{barr2021role}.  

Figure \ref{fig:concVsEandEV} and the linear model given in Table \ref{tab:linearmodel} show that the concentration is dependent on GB energy, $\gamma$, and GB excess volume, $V_\mathrm{exc}$, for GBs sampled over the entire 5D space. However, the linear model for $c_\mathrm{GB}$ has almost no dependence on the disorientation angle, $\theta_\mathrm{dis}$. This lack of a relationship with the disorientation angle as well as the rugged landscape in 5D space shown in Figure \ref{fig:cakePlots} make it difficult to understand how to optimize GB segregation energy characteristics through texture control or grain boundary engineering. This remains a challenge to be solved.

\section{Conclusions}

In this work, we examined grain boundary segregation energy spectra in \numGBs GBs that comprehensively sample the 5D space of crystallographic character. This included calculating segregation energies for more than \numAtoms possible sites. 

Verification of the dataset involved determining how to identify bulk vs.\ GB atoms. While aCNA is often used for bulk determination in GBs, the number of non-negligible segregation energies that were excluded by aCNA categorization caused us to consider CSP as an alternative, with a cutoff of $0.1$ for bulk determination. The differences between these two methods are illustrated in Figures \ref{fig:tableOfGBpics} and \ref{fig:CSPvsCNAbulk} and discussed further in Supplemental Section \ref{sec:CNAvsCSP}. The resulting segregation energy spectra from both methods were also compared in Figures \ref{fig:sigma5}, \ref{fig:catDataHist} and \ref{fig:SymTwistGBs}b. It was determined that they give similar answers despite the reduced number of atoms included by aCNA, as illustrated in a comparison of $c_\mathrm{GB}$ values calculated by both CSP and aCNA shown in Supplemental Figure \ref{fig:concCSPvsCNA}. Verification also involved removing invalid segregation energy calculations that didn't converge to $0$ eV at large distances from the GB, since they don't match the expected behavior  \cite{scheiber2015ab, jin2014study}, and GBs with unreasonably low segregation energies. The complete list of GBs excluded from analysis in this work is given in Supplemental Table \ref{tab:excludedGBs}.

Validation involved comparing the computed GB segregation energies to similar bicrystal and polycrystal studies. In Figure \ref{fig:sigma5} it was shown that a subset of the data produced in this work is similar to the work of Huber et al.\ in \cite{huber2018machine}. The aggregated spectrum of segregation energies in the Homer bicrystal GB dataset is also similar to the segregation energy spectrum in polycrystals obtained by Wagih et al.\ in \cite{wagih2020learning}, as shown in Figure \ref{fig:catDataHist} and Table \ref{tab:catDataHist}. Both validation comparisons are favorable, but some minor differences between the polycrystal and bicrystal spectra raise a number of questions, posed in Section \ref{sec:polycrystalComparison}, that are worth resolving and that could impact the quality of an aggregate segregation energy spectrum.

Several insights arose from different methods of analysis.
Figure \ref{fig:segVsGamma} shows that as GB energy increases, Co segregation in Al GBs becomes more favorable. This is supported by the increase of $c_\mathrm{GB}$ with GB energy shown in Figure \ref{fig:histOfFracSeg}b, Figure \ref{fig:concVsEandEV}, and Supplemental Figure \ref{fig:concVsGamma}. Additionally, 
all of the GBs have higher $c_\mathrm{GB}$ than $c_\mathrm{bulk}$, except the [111] symmetric twist perfect twin GB.
Figure \ref{fig:histOfFracSeg}b shows that most GBs have a preference for segregation, evidenced by their proximity to the ``all segregating'' corner of the plot. However, there are some some interesting GBs that are outliers, many of which have [100], [110], and [111] disorientation axes. The most extreme outlier is the [111] twin GB that is located at ``all anti-segregating''. In addition, the temperature dependence of $c_\mathrm{GB}$ was demonstrated in Figure \ref{fig:SymTwistGBs}c, which shows that $c_\mathrm{GB}$ drops dramatically at higher temperatures. 

In general, it was found that $c_\mathrm{GB}$ has smooth variation across the 5D space of crystallographic character (see Figure \ref{fig:cakePlots} and Supplemental Figure \ref{fig:concSymmetric}). Additional examination showed that $c_\mathrm{GB}$ is correlated with GB energy and GB excess volume (see Figure \ref{fig:concVsEandEV} and Table \ref{tab:linearmodel}), but is not correlated with disorientation angle. $c_\mathrm{GB}$ does not have an obvious functional form in 5D crystallographic space, which is shown by the rugged landscapes in Figure \ref{fig:cakePlots} and the lack of obvious trends in Supplemental Figure \ref{fig:concSymmetric}. As segregation is an active research topic for the development of advanced materials, the dataset containing all of the per GB quantities computed in this work, including the $c_\mathrm{GB}$ values and a histogram of each GB's segregation spectrum, is available for download \cite{CoAlSegDataset}.

\section*{Acknowledgments}
This work was supported by the U.S.\ National Science Foundation (NSF) under Award \#DMR-1817321. 

\bibliographystyle{elsarticle-num}
\bibliography{ResearchPaperBib}



\clearpage
\setcounter{section}{0}
\setcounter{page}{1}
\setcounter{figure}{0}
\setcounter{equation}{0}
\setcounter{table}{0}
\renewcommand{\thesection}{S\arabic{section}}
\renewcommand{\thepage}{S\arabic{page}}
\renewcommand{\thetable}{S\arabic{table}}
\renewcommand{\thefigure}{S\arabic{figure}}

\setcounter{affn}{0}
\resetTitleCounters

\makeatletter
\let\@title\@empty
\makeatother

\title{\section*{Supplementary Materials}\documenttitle}

\makeatletter
\renewenvironment{abstract}{\global\setbox\absbox=\vbox\bgroup
  \hsize=\textwidth\def\baselinestretch{1}%
 \noindent\unskip}
 {\egroup}

\def\ps@pprintTitle{%
     \let\@oddhead\@empty
     \let\@evenhead\@empty
     \def\@oddfoot{\footnotesize\itshape
        Supplementary Materials for \ifx\@journal\@empty Elsevier
       \else\@journal\fi\hfill\today}%
     \let\@evenfoot\@oddfoot}
\makeatother

\begin{abstract}
This document contains supplementary information for the main article.
\end{abstract}
\maketitle



\section{Comparison of bulk atom selection by aCNA and CSP} \label{sec:CNAvsCSP}

To contrast the differences between bulk atom selection by aCNA and CSP, we examine a few different items. First, we note that 90\% of the atoms classified as bulk FCC by aCNA have CSP values less than 0.1, as illustrated in Supplemental Figure \ref{fig:CSPhistogram}. However, the remaining 10\% of atoms classified as bulk FCC by aCNA can have CSP values as high as 5, indicating the tendency for aCNA to tolerate some noise in the classification of environments as as the FCC structure. Second, we present a contingency table in Figure \ref{fig:CSPvsCNA_contingency} showing that of the \numAtoms atoms simulated, only 17\% (11.5 million) were classified by aCNA as GB atoms, compared to the 26\% (18 million) atoms classified as GB atoms by CSP. This also results in a larger bulk atom population for aCNA than CSP. The segregation energy spectrum for these bulk atoms by both classification schemes is illustrated in Figure \ref{fig:CSPvsCNAbulk}. The aCNA spectrum is slightly wider than the CSP spectrum because atoms with larger structural distortions and therefore a wider distribution of segregation energies are categorized as FCC by the aCNA classifier. In short, the aCNA and CSP give fairly similar results, with aCNA being more restrictive in its selection of GB atoms and therefore leaving out some atoms with non-negligble segregation energies. On the other hand, classification of bulk atoms by CSP $\leq 0.1$ results in a more generous selection of GB atoms that capture the segregation energy deviations left out by aCNA. But, this comes at the cost of having more GB atoms with very small segregation energies. The effect of these near-negligible segregation energies on the $c_\mathrm{GB}$ in the CSP spectra is shown in Supplemental Figure \ref{fig:concCSPvsCNA}; CSP predicts lower $c_\mathrm{GB}$ values, however any trends in $c_\mathrm{GB}$ appear to be the same whether CSP or aCNA were used due to the positive correlation between the two methods. Perhaps a more restrictive CSP value might bridge this difference better but we do not attempt to fine tune the CSP value in this work.

\begin{figure}[h]
    \centering
    \includegraphics[width=\columnwidth]{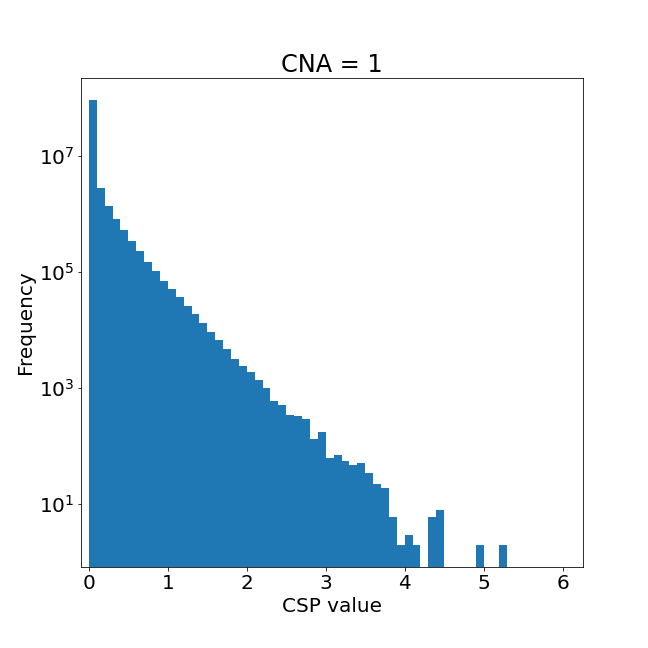}
    \caption{Logarithmic scale of the distribution of centrosymmetry parameters (CSP) for FCC atoms as determined by aCNA. The bin size is $0.1$. Only 7\% of FCC atoms determined by aCNA have a CSP greater than $0.1$. 
    }
    \label{fig:CSPhistogram}
\end{figure}

\begin{figure}[h]
\begin{centering}
\begin{tabular}{ccccc}
                                          &                           & \multicolumn{2}{c}{CSP}                                                                                                                           &                                                                         \\ \cline{3-4}
                                          & \multicolumn{1}{c|}{}     & \multicolumn{1}{c|}{GB}                                                 & \multicolumn{1}{c|}{Bulk}                                               &                                                                         \\ \cline{2-5} 
\multicolumn{1}{c|}{\multirow{2}{*}{aCNA}} & \multicolumn{1}{c|}{GB}   & \multicolumn{1}{c|}{11.5M}                                                   & \multicolumn{1}{c|}{0}                                                   & \multicolumn{1}{c|}{\begin{tabular}[c]{@{}c@{}}11.5M\\ 17\%\end{tabular}} \\ \cline{2-5} 
\multicolumn{1}{c|}{}                     & \multicolumn{1}{c|}{Bulk} & \multicolumn{1}{c|}{6.5M}                                                   & \multicolumn{1}{c|}{51M}                                                   & \multicolumn{1}{c|}{\begin{tabular}[c]{@{}c@{}}57.5M\\ 83\%\end{tabular}} \\ \cline{2-5} 
                                          & \multicolumn{1}{c|}{}     & \multicolumn{1}{c|}{\begin{tabular}[c]{@{}c@{}}18M\\ 26\%\end{tabular}} & \multicolumn{1}{c|}{\begin{tabular}[c]{@{}c@{}}51M\\ 74\%\end{tabular}} &                                                                         \\ \cline{3-4}
\end{tabular}
\caption{Contingency table comparing CSP and aCNA bulk atom determination. All atoms determined by aCNA to be bulk are also classified as bulk by CSP, although aCNA includes 6.5 million additional atoms in the bulk category. }
\label{fig:CSPvsCNA_contingency}
\end{centering}
\end{figure} 

\begin{figure}[h]
    \centering
    \includegraphics[width=\columnwidth]{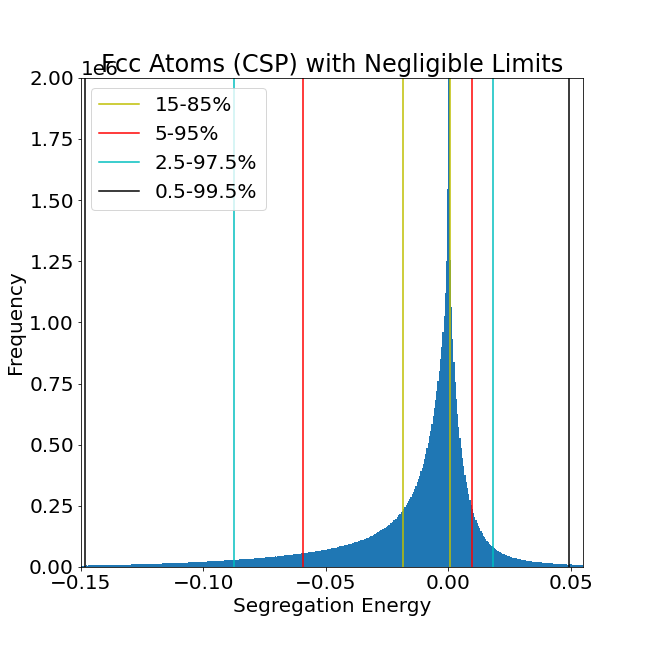}
    \caption{
    Distribution of segregation energies for FCC atoms as determined by CSP. Different intervals are shown. This work uses the 95\% interval for the limits of the ``negligible segregation'' category (\mbox{$-0.040$ eV $\leq E_{\mathrm{seg}}^{\mathrm{Co}_i} < +0.0115$ eV}). }
    \label{fig:neglLimitsFCC}
\end{figure}

\begin{figure}[h]
    \centering
    \includegraphics[width=\columnwidth]{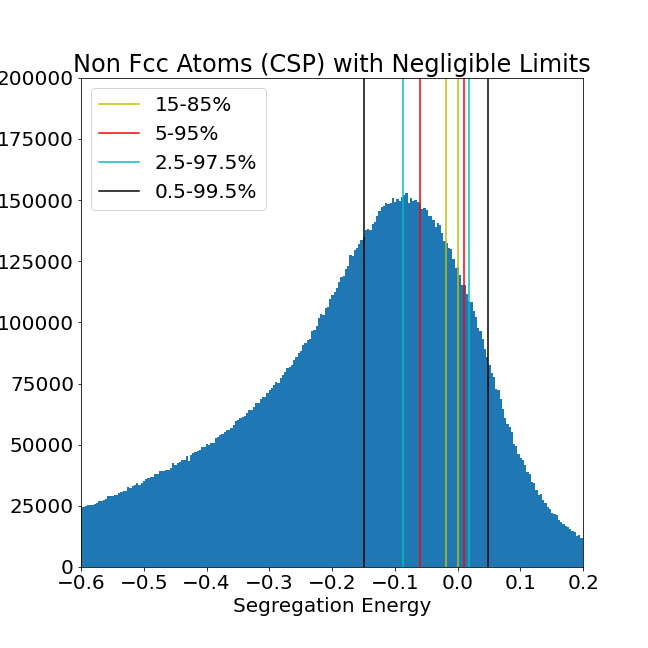}
    \caption{
    Distribution of segregation energies for non-FCC atoms as determined by CSP. Different intervals of the distribution of segregation energies for FCC atoms as determined by CSP are shown. This work uses the 95\% interval for the limits of the ``negligible segregation'' category (\mbox{$-0.040$ eV $\leq E_{\mathrm{seg}}^{\mathrm{Co}_i} < +0.0115$ eV}).}
    \label{fig:neglLimitsNonFCC}
\end{figure}

\begin{figure}[h]
    \centering
    \includegraphics[width=\columnwidth]{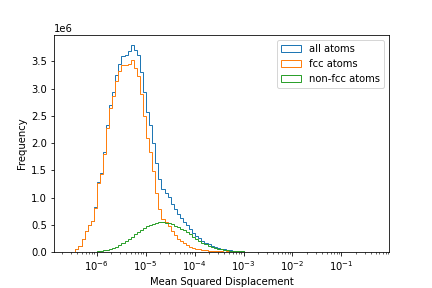}
    \caption{Mean Squared Displacement distribution of all atoms (blue) all FCC atoms as determined by aCNA (orange), and all non-FCC GB atoms (green). All of the distributions are approximately log normal.}
    \label{fig:MSDhist}
\end{figure}

\begin{figure}[h]
    \centering
    \includegraphics[width=\columnwidth]{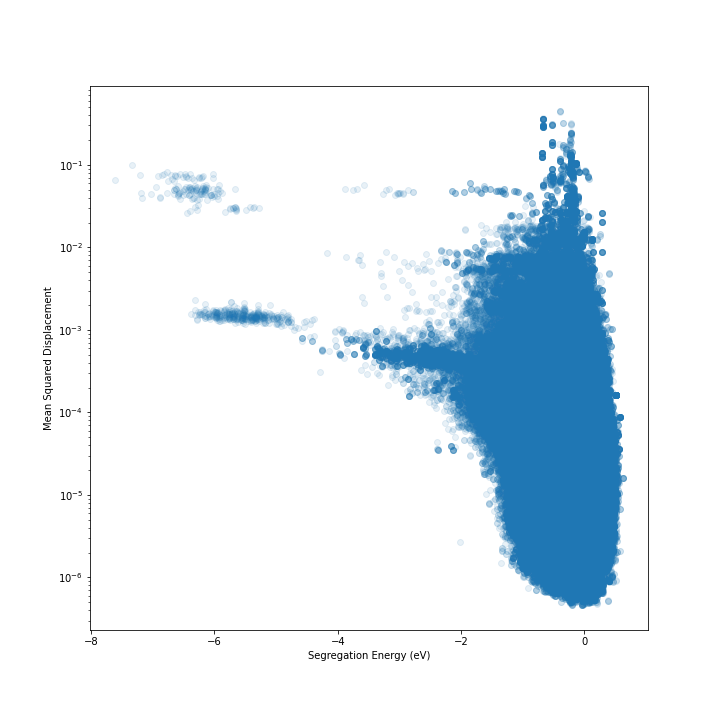}
    \caption{Scatter plot of mean squared displacement (MSD) vs.\ segregation energy ($E_{\mathrm{seg}}^{\mathrm{Co}_i}$) for all non-FCC atoms. This scatter plot was used to decide on a (later discarded) ``high-MSD'' cutoff for atoms to remove from analysis, noting that the approximately skew-normal distribution of segregation energies was interrupted mostly by atoms with MSD $> 10^{-4}$, which was decided somewhat arbitrarily to be the ``high-MSD'' cutoff value considered in this work. }
    \label{fig:MSDvsSegEng}
\end{figure}

\begin{figure}[h]
    \centering
    \includegraphics[width=\columnwidth]{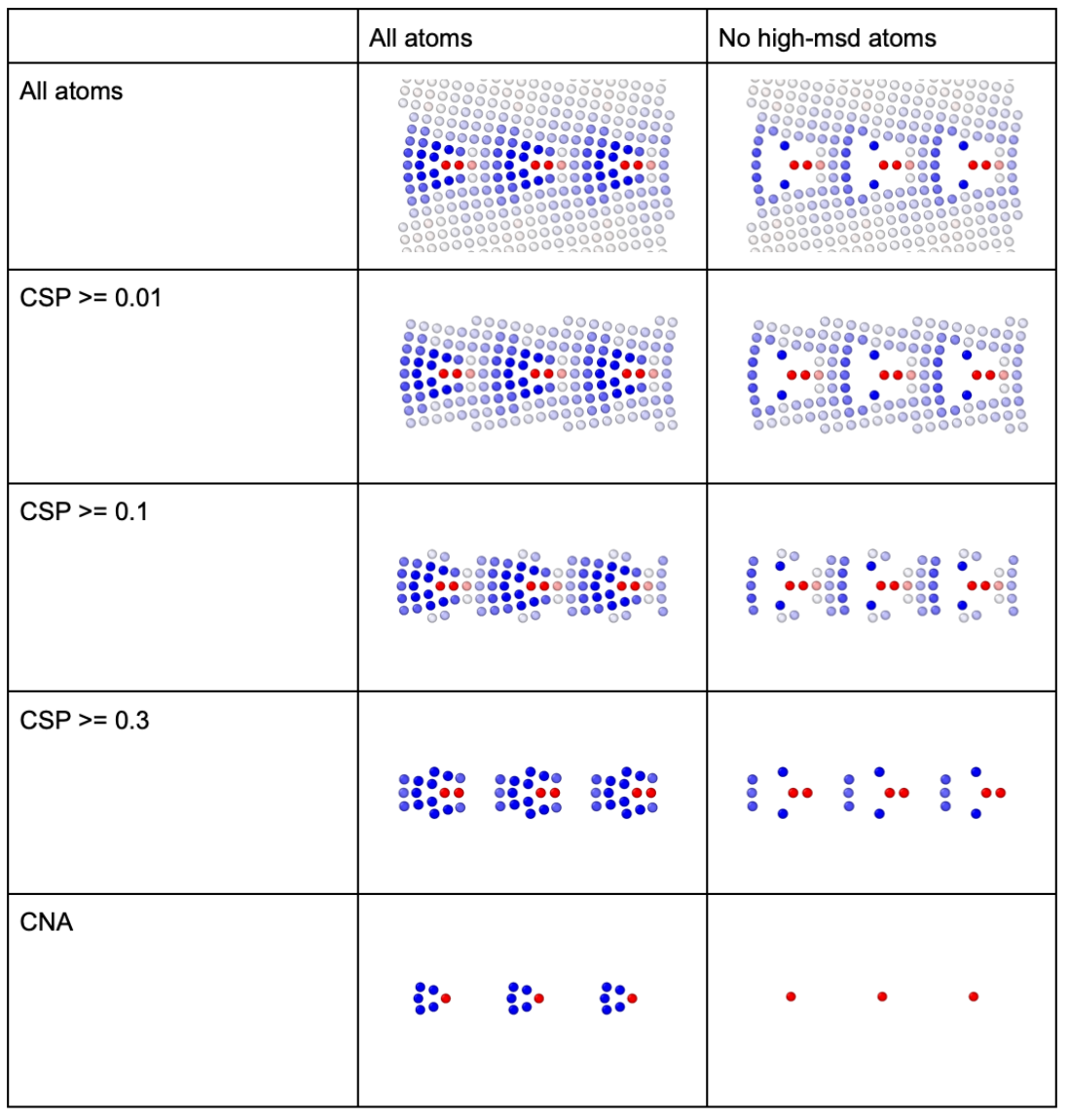} 
    \caption{A [100] symmetric tilt GB with an array of edge dislocations with all atoms, and bulk atoms cut out using 4 different techniques: \mbox{CSP $\leq 0.3$}, \mbox{CSP $\leq 0.1$}, CSP \mbox{$\leq 0.01$}, and aCNA. The right side also cuts out atoms according to the high-MSD cutoff of $10^{-4}$. Red atoms have positive segregation energies, blue have negative, and grey have negligible, according to the colorbar shown. On the right, using the combination of high-MSD and CNA, only one atom per structural unit \cite{Frost:1982dv, Sutton:1983vi, Balluffi:1984kk, Rittner:1996vf, Tschopp:2007hr, Spearot:2008bq, Han:2017io} in the GB remains to analyze, giving the impression that the entire GB is unfavorable to solute segregation (red coloring corresponds to anti-segregation), when other removal methods show that the GB contains many other sites favorable to solute segregation (blue colored atom sites). Such aggressive atom removal via aCNA and MSD caused the authors to consider other methods of atom removal, discussed in Sections \ref{sec:bulkSeg} and \ref{sec:restructuring}.}
    \label{fig:tableOfGBpicsMSD}
\end{figure}

\begin{figure}[h]
    \centering
    \includegraphics[width=\columnwidth]{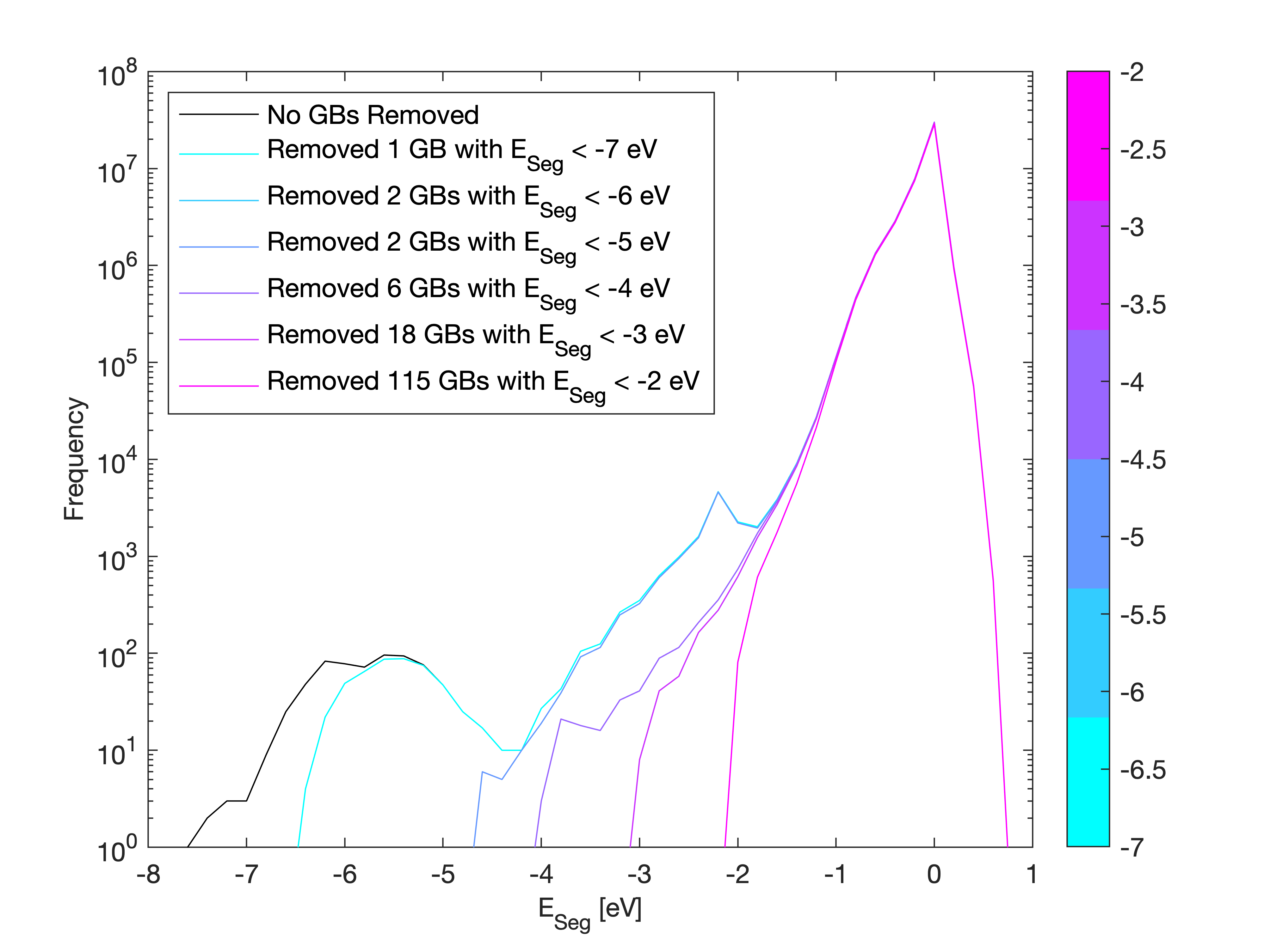}
    \caption{Spectrum of segregation energies for the entire Homer GB dataset \cite{Homer:2022:AlGBdataset}, with all atoms from GBs below several different cutoffs removed. A high concentration of extreme segregation energy values were found in several GBs. The authors chose to remove 18 GBs.}
    \label{fig:NgbsRemoved}
\end{figure}

\begin{figure}[h]
    \centering
    \includegraphics[width=\columnwidth]{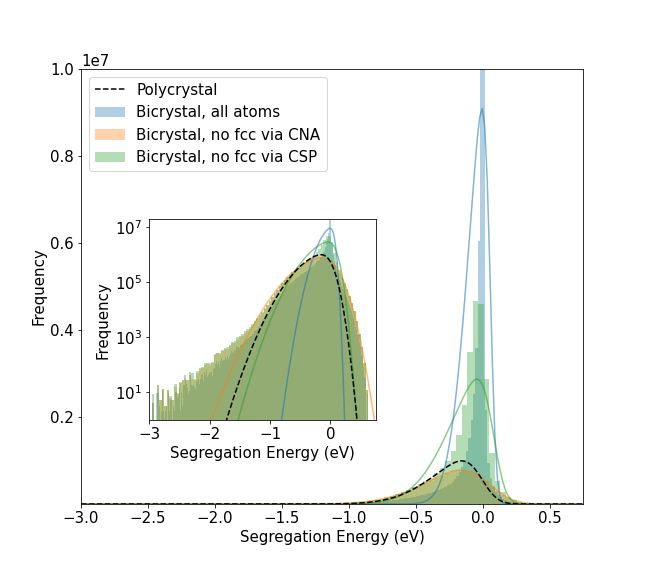}
    \caption{The spectrum of segregation energies aggregated from all GBs simulated in this work, including bulk atoms (blue), with all bulk atoms removed via aCNA (orange), and via CSP (green). The dotted line is the polycrystal spectrum from Wagih et al.\ in \cite{wagih2020learning}. Compare to Figure \ref{fig:catDataHist}}. 
    \label{fig:catDataHistwithALL}
\end{figure}

\begin{figure}[h]
    \centering
     \includegraphics[width=\columnwidth]{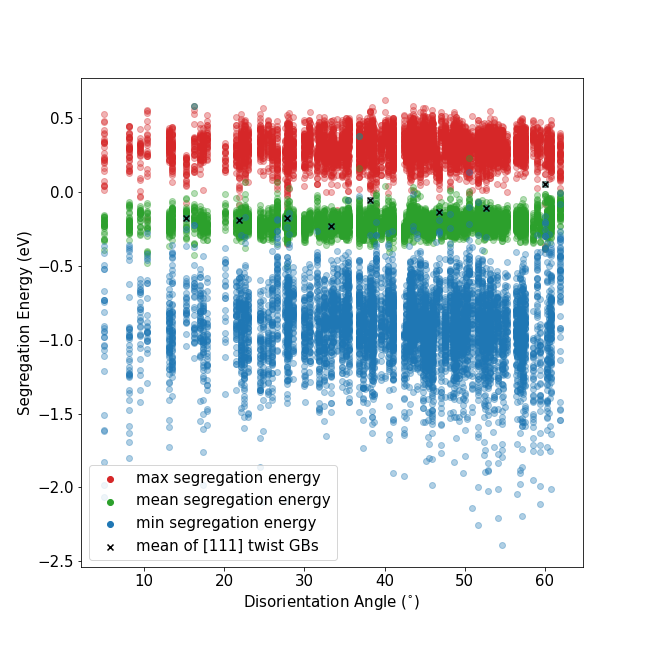}
    \caption{ Minimum (blue), maximum (red), and mean (green) of segregation energy $E_{\mathrm{seg}}^{\mathrm{Co}_i}$ vs.\ their disorientation angle. [111] symmetric twist GBs marked by black ``x''s.} 
    \label{fig:segVsDis}
\end{figure}

\begin{figure}[h]
    \centering
    \includegraphics[width=\columnwidth]{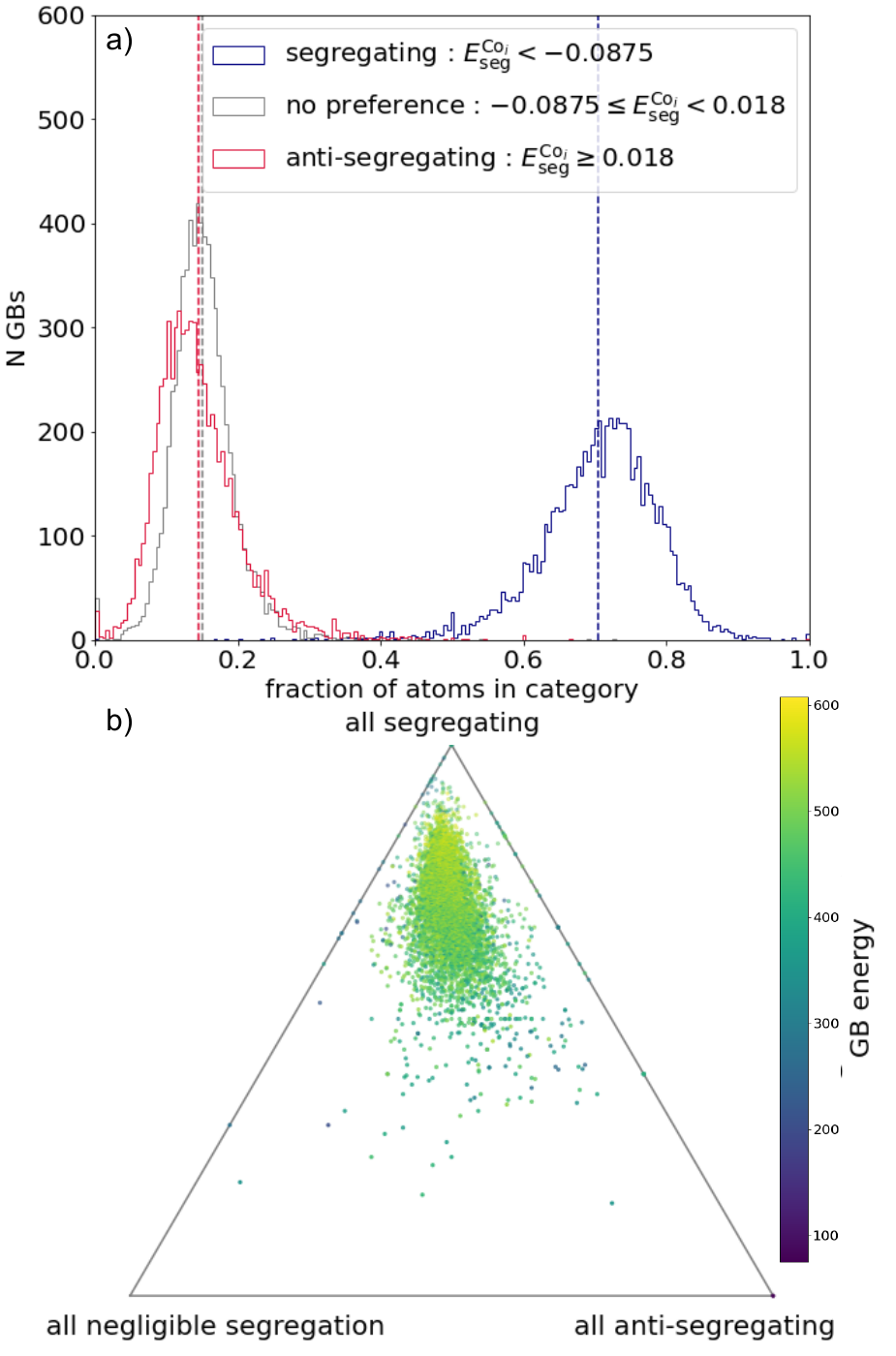}
    \caption{Ternary Plot for CNA.}
    \label{fig:ternaryPlotCNA}
\end{figure}

\begin{figure}[h]
    \centering
    \includegraphics[width=\columnwidth]{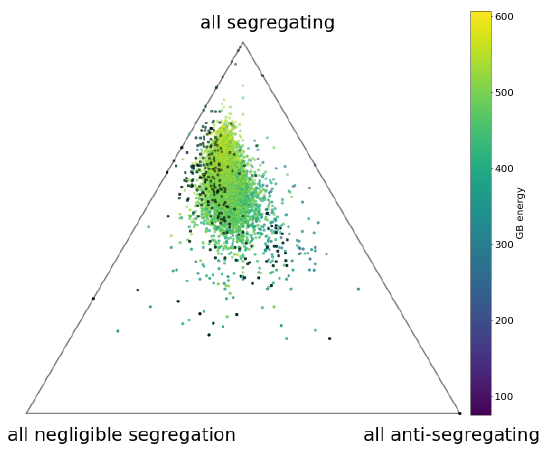}
    \caption{Ternary Plot for CSP with the [111] disorientation axis GBs shown in black.}
    \label{fig:ternaryPlot111}
\end{figure}

\begin{figure}[h]
    \centering
    \includegraphics[width=\columnwidth]{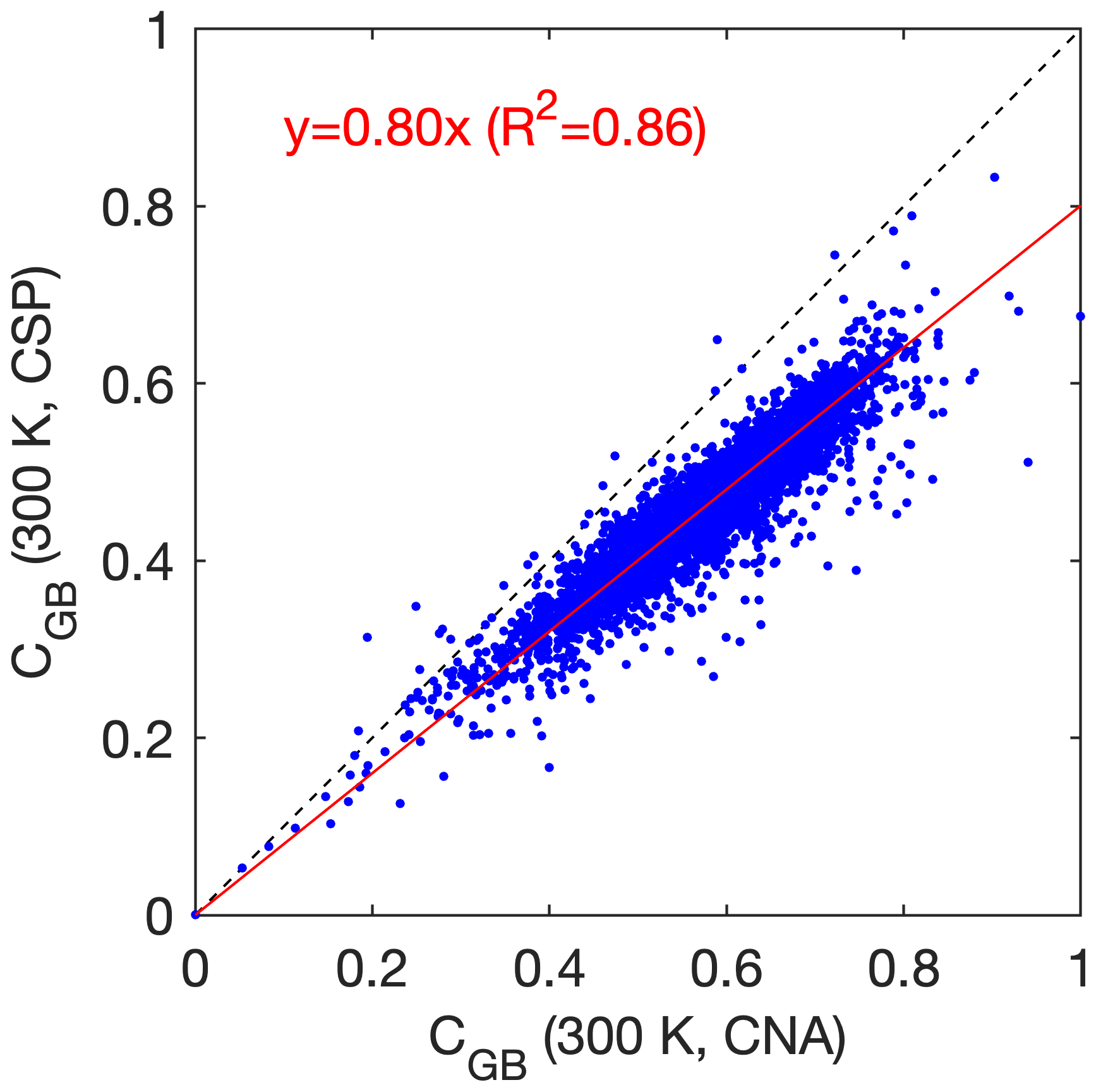}
    \caption{Correlation plot between $c_\mathrm{GB}$ as calculated using GB atoms determined by aCNA (x-axis) and CSP (y-axis) for all \numGBs GBs. CSP bulk determination includes more negligibly segregating atom sites, and therefore has a lower $c_\mathrm{GB}$ than the corresponding value for aCNA bulk determination, with a correlation of $c_\mathrm{GB}^\mathrm{CSP} = .8 c_\mathrm{GB}^\mathrm{aCNA}$.}
    \label{fig:concCSPvsCNA}
\end{figure}

\begin{figure}[h]
    \centering
    \includegraphics[width=\columnwidth]{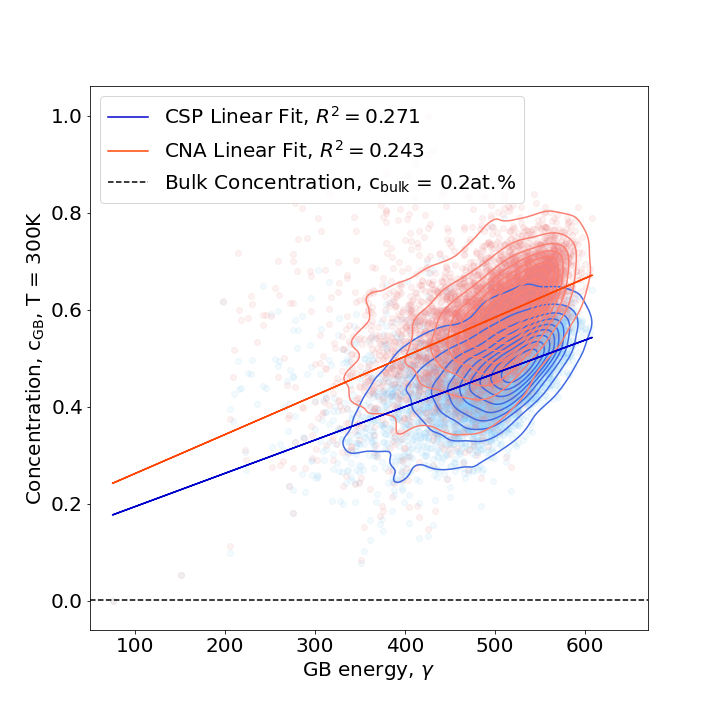}
    \caption{
    Concentration of solute at the GB, $c_{\mathrm{GB}}$ (shown in fractions, not at\%), vs.\ GB energy, $\gamma$, for all GBs in the Homer dataset \cite{Homer:2022:AlGBdataset} at  \mbox{T = 300 K}, for aCNA (red) and CSP (blue) bulk atom determination. Density contours overlaid, with linear fit line with $R^2$ values indicated in the legend. All GBs except the [111] symmetric twist perfect twin have a higher concentration, $c_{\mathrm{GB}}$, than \mbox{$c_{\mathrm{bulk}}$=0.2at\%}, shown with a black dotted line. } 
    \label{fig:concVsGamma}
\end{figure}

\begin{figure*}[t]
    \centering
    \includegraphics[width=.75\textwidth]{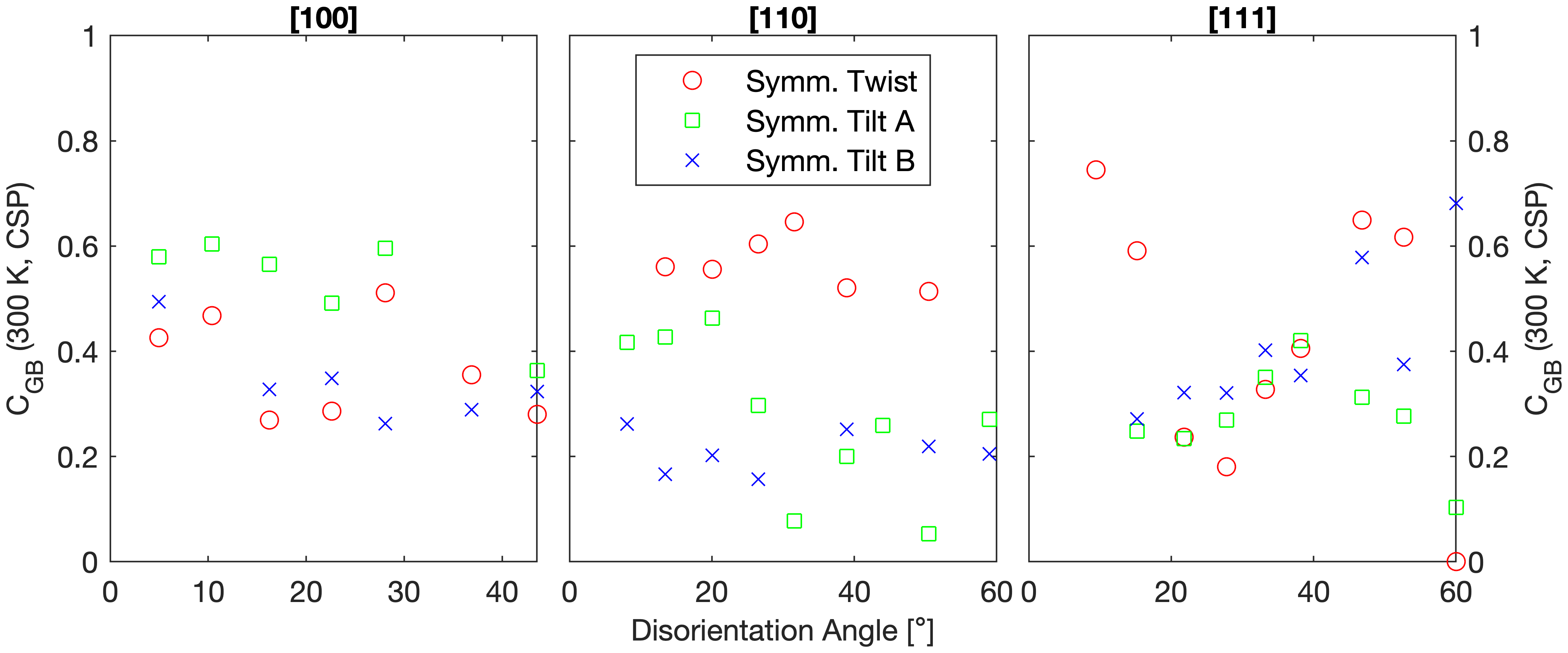}
    \caption{$c_\mathrm{GB}$ (CSP) vs.\ disorientation angle ($^\circ$) for 9 special types of GBs: the symmetric twist GBs (red) and the symmetric tilt GBs (green + blue) for the [100], [110], and [111] disorientation axes. None of these subsets demonstrate the general trends very well, but the [111] symmetric twist GBs are noted in the main body of this paper for illustration purposes. (Note: given are the disorientation angles about the disorientation axes but for the symmetric tilt GBs, their tilt angle is a derived quantity.) }
    \label{fig:concSymmetric}
\end{figure*}

\begin{figure*}[t]
    \centering
    \includegraphics[width=.9\textwidth]{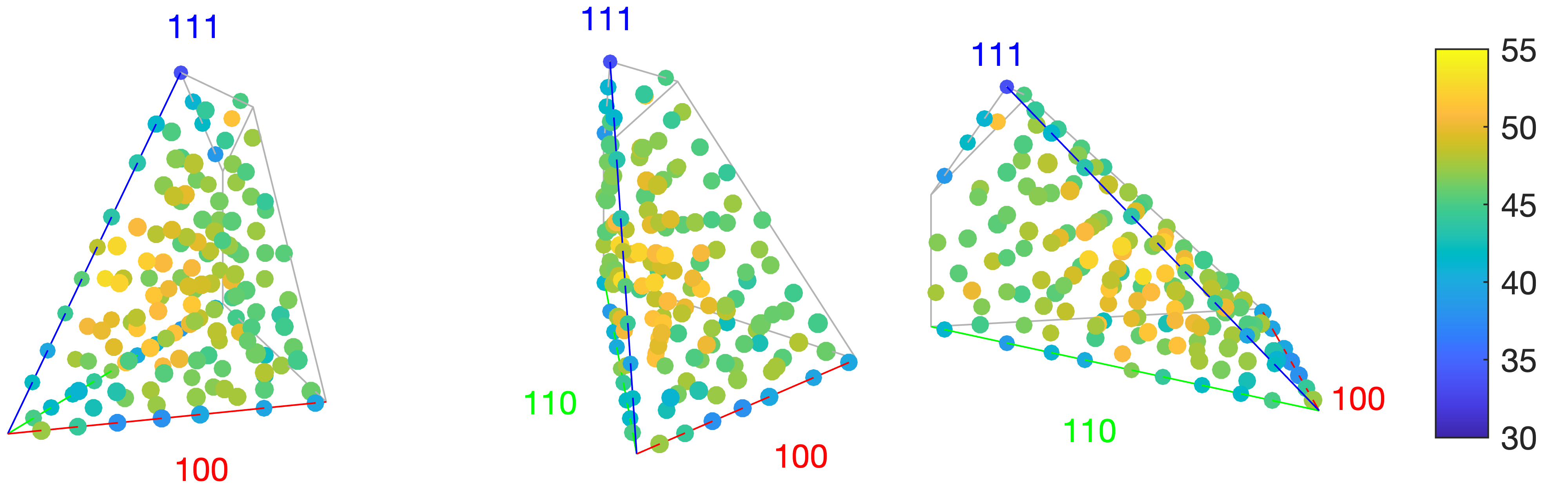}
    \caption{Multiple views of $c_\mathrm{GB}$ ($100\times$at\%) in Rodriguez-Frank space. The following disorientation axes are found along the corresponding (color)ed lines: [100] (red), [110] (green), [111] (blue).}
    \label{fig:concRFspace}
\end{figure*}

\begin{figure}[h]
    \centering
    \includegraphics[width=\columnwidth]{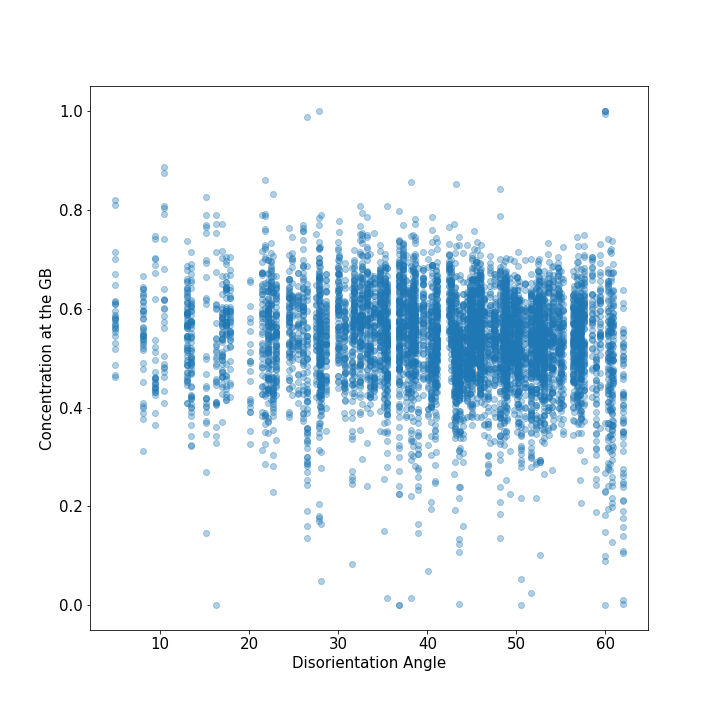}
    \caption{Concentration of solute at the GB, $c_\mathrm{GB}$, as a function of disorientation angle.}
    \label{fig:concVsDis}
\end{figure}

\clearpage
\onecolumn

\clearpage
\begin{table}[h]
    \caption{GBs excluded from the Al Homer dataset \cite{Homer:2022:AlGBdataset}. The CSL value, disorientation (dis.) angle and axis, and boundary plane normals from both sides of the grain $\mathrm{N}_\mathrm{A}$ / $\mathrm{N}_\mathrm{B}$ are provided, along with the reason the GB is excluded from the analysis.}
    \label{tab:GBsexcluded}
\begin{center}
\begin{supertabular}{cccc}
        \hline
        CSL & Dis. angle and axis & $\mathrm{N}_\mathrm{A}$ / $\mathrm{N}_\mathrm{B}$ & Reason excluded \\\hline 

$\Sigma145h$ & $56.5^{\circ} \left[6\,5\,2\right]$ & $\left(29\,165\,153\right)/\left(\overline{5}\,\overline{45}\,\overline{3}\right)$ & Refill Issues \\ 
$\Sigma265a$ & $5.0^{\circ} \left[1\,0\,0\right]$ & $\left(0\,\overline{37}\,56\right)/\left(0\,32\,\overline{59}\right)$ & Refill Issues \\
$\Sigma265a$ & $5.0^{\circ} \left[1\,0\,0\right]$ & $\left(265\,\overline{293}\,379\right)/\left(\overline{265}\,259\,\overline{403}\right)$ & Refill Issues \\
$\Sigma265a$ & $5.0^{\circ} \left[1\,0\,0\right]$ & $\left(265\,\overline{239}\,727\right)/\left(\overline{53}\,35\,\overline{149}\right)$ & Refill Issues \\
$\Sigma265a$ & $5.0^{\circ} \left[1\,0\,0\right]$ & $\left(265\,\overline{215}\,1235\right)/\left(\overline{265}\,107\,\overline{1249}\right)$ & Refill Issues \\
$\Sigma303a$ & $24.8^{\circ} \left[3\,2\,1\right]$ & $\left(\overline{58}\,85\,307\right)/\left(106\,\overline{191}\,\overline{239}\right)$ & Refill Issues \\
$\Sigma441o$ & $49.2^{\circ} \left[11\,4\,4\right]$ & $\left(269\,\overline{212}\,\overline{638}\right)/\left(\overline{269}\,638\,212\right)$ & Refill Issues \\ 
$\Sigma647q$ & $60.3^{\circ} \left[9\,9\,1\right]$ & $\left(141\,\overline{185}\,396\right)/\left(185\,\overline{141}\,\overline{396}\right)$ & Refill Issues \\ 
$\Sigma943l$ & $38.4^{\circ} \left[10\,1\,1\right]$ & $\left(117\,76\,\overline{303}\right)/\left(\overline{135}\,132\,275\right)$ & Refill Issues \\ 
$\Sigma999ae$ & $58.5^{\circ} \left[16\,11\,10\right]$ & $\left(721\,\overline{61}\,\overline{587}\right)/\left(\overline{685}\,625\,\overline{91}\right)$ & Refill Issues \\ 

$\Sigma131e$ & $60.3^{\circ} \left[5\,5\,4\right]$ & $\left(21\,\overline{1147}\,1211\right)/\left(1157\,\overline{39}\,\overline{1201}\right)$ & Conv. Issues \\ 
$\Sigma131e$ & $60.3^{\circ} \left[5\,5\,4\right]$ & $\left(\overline{97}\,\overline{703}\,869\right)/\left(833\,\overline{137}\,\overline{739}\right)$ & Conv. Issues \\ 
$\Sigma265a$ & $5.0^{\circ} \left[1\,0\,0\right]$ & $\left(265\,\overline{63}\,389\right)/\left(\overline{265}\,29\,\overline{393}\right)$ & Conv. Issues \\ 
$\Sigma549h$ & $32.4^{\circ} \left[13\,1\,1\right]$ & $\left(139\,\overline{5}\,\overline{155}\right)/\left(\overline{143}\,91\,121\right)$ & Conv. Issues \\ 

$\Sigma55b$ & $38.6^{\circ} \left[2\,1\,1\right]$ & $\left(\overline{39}\,38\,\overline{70}\right)/\left(11\,0\,88\right)$ & Low $E_{seg}$ Values \\ 
$\Sigma55c$ & $57.6^{\circ} \left[5\,5\,1\right]$ & $\left(67\,\overline{26}\,70\right)/\left(\overline{4}\,\overline{32}\,\overline{95}\right)$ & Low $E_{seg}$ Values \\ 
$\Sigma55c$ & $57.6^{\circ} \left[5\,5\,1\right]$ & $\left(109\,\overline{117}\,95\right)/\left(9\,17\,\overline{185}\right)$ & Low $E_{seg}$ Values \\ 
$\Sigma73b$ & $13.4^{\circ} \left[1\,1\,0\right]$ & $\left(192\,\overline{119}\,114\right)/\left(\overline{169}\,96\,\overline{162}\right)$ & Low $E_{seg}$ Values \\ 
$\Sigma73b$ & $13.4^{\circ} \left[1\,1\,0\right]$ & $\left(235\,\overline{89}\,119\right)/\left(\overline{211}\,65\,\overline{169}\right)$ & Low $E_{seg}$ Values \\ 
$\Sigma73e$ & $48.9^{\circ} \left[4\,3\,0\right]$ & $\left(\overline{507}\,311\,331\right)/\left(543\,\overline{359}\,199\right)$ & Low $E_{seg}$ Values \\ 
$\Sigma83d$ & $57.2^{\circ} \left[3\,3\,1\right]$ & $\left(259\,\overline{681}\,685\right)/\left(429\,65\,\overline{901}\right)$ & Low $E_{seg}$ Values \\ 
$\Sigma145h$ & $56.5^{\circ} \left[6\,5\,2\right]$ & $\left(609\,\overline{515}\,\overline{467}\right)/\left(\overline{477}\,695\,\overline{379}\right)$ & Low $E_{seg}$ Values \\ 
$\Sigma145h$ & $56.5^{\circ} \left[6\,5\,2\right]$ & $\left(29\,\overline{55}\,123\right)/\left(53\,\overline{45}\,\overline{119}\right)$ & Low $E_{seg}$ Values \\ 
$\Sigma197g$ & $54.1^{\circ} \left[9\,9\,1\right]$ & $\left(291\,\overline{11}\,41\right)/\left(\overline{205}\,\overline{57}\,\overline{203}\right)$ & Low $E_{seg}$ Values \\ 
$\Sigma197g$ & $54.1^{\circ} \left[9\,9\,1\right]$ & $\left(248\,\overline{335}\,586\right)/\left(215\,\overline{118}\,\overline{676}\right)$ & Low $E_{seg}$ Values \\ 
$\Sigma197g$ & $54.1^{\circ} \left[9\,9\,1\right]$ & $\left(111\,\overline{451}\,499\right)/\left(307\,45\,\overline{607}\right)$ & Low $E_{seg}$ Values \\ 
$\Sigma201a$ & $8.1^{\circ} \left[1\,1\,0\right]$ & $\left(241\,\overline{40}\,197\right)/\left(\overline{220}\,19\,\overline{223}\right)$ & Low $E_{seg}$ Values \\ 
$\Sigma201a$ & $8.1^{\circ} \left[1\,1\,0\right]$ & $\left(779\,\overline{377}\,103\right)/\left(\overline{763}\,361\,\overline{217}\right)$ & Low $E_{seg}$ Values \\ 
$\Sigma215e$ & $45.9^{\circ} \left[9\,7\,1\right]$ & $\left(\overline{189}\,323\,\overline{345}\right)/\left(\overline{9}\,\overline{9}\,101\right)$ & Low $E_{seg}$ Values \\ 
$\Sigma265a$ & $5.0^{\circ} \left[1\,0\,0\right]$ & $\left(265\,\overline{689}\,1007\right)/\left(\overline{265}\,599\,\overline{1063}\right)$ & Low $E_{seg}$ Values \\ 
$\Sigma283f$ & $44.7^{\circ} \left[9\,1\,0\right]$ & $\left(\overline{515}\,107\,273\right)/\left(531\,\overline{251}\,\overline{79}\right)$ & Low $E_{seg}$ Values \\ 
$\Sigma303a$ & $24.8^{\circ} \left[3\,2\,1\right]$ & $\left(\overline{46}\,151\,139\right)/\left(50\,\overline{193}\,\overline{67}\right)$ & Low $E_{seg}$ Values \\ 

        \hline
    
\end{supertabular}\label{tab:excludedGBs}
\end{center}

\end{table}

\end{document}